\documentclass[preprint,11pt]{elsarticle}

\usepackage{amssymb}
\usepackage{amsmath}
\usepackage{bm}
\usepackage{siunitx}
\usepackage[a4paper,total={7in,10in}]{geometry}
\usepackage{upgreek}
\usepackage{booktabs}
\usepackage{multirow}
\usepackage{makecell}
\usepackage{threeparttable}
\usepackage{longtable}
\usepackage[section]{placeins}

\journal{International Journal of Mechanical Science}

\begin{document}

\begin{frontmatter}

\title{An RBC-MsUQ Framework for Red Blood Cell Morpho-Mechanics}

\author[inst1]{Shuo Wang}
\author[inst2]{Lei Ma}
\author[inst2]{Ling Guo\corref{cor1}}
\ead{lguo@shnu.edu.cn}
\author[inst1]{Xuejin Li\corref{cor1}}
\ead{xuejin_li@zju.edu.cn}
\cortext[cor1]{Corresponding author}
\author[inst3]{Tao Zhou}

\affiliation[inst1]{organization={State Key Laboratory of Fluid Power and Mechatronic Systems, Department of Engineering Mechanics and Center for X-Mechanics},%Department and Organization
            addressline={Zhejiang University}, 
            city={Hangzhou},
            postcode={310027},
            country={China}}

\affiliation[inst2]{organization={Mathematics and Science College},%Department and Organization
            addressline={Shanghai Normal University}, 
            city={Shanghai},
            postcode={200234},
            country={China}}

\affiliation[inst3]{organization={Institute of Computational Mathematics and Scientific/Engineering Computing, Academy of Mathematics and Systems Sciences},%Department and Organization
            addressline={Chinese Academy of Sciences}, 
            city={Beijing},
            postcode={100190},
            country={China}}       

%% Abstract
\begin{abstract}
%genetic features of biological cells
%Red blood cells (RBCs) are critical components of the human circulatory system, facilitating oxygen and nutrient transport through their unique morphology and mechanical properties. 
Characterizing the morpho-mechanical properties of red blood cells (RBCs) is crucial for understanding microvascular transport mechanisms and cellular pathophysiological processes, yet current computational models are constrained by multi-source uncertainties including cross-platform experimental discrepancies and parameter identification stochasticity. We present RBC-MsUQ, a novel multi-stage uncertainty quantification framework tailored for RBCs. It integrates hierarchical Bayesian inference with diverse experimental datasets, establishing prior distributions for RBC parameters via microscopic simulations and literature-derived data. A dynamic annealing technique defines stress-free baselines, while deep neural network surrogates, optimized through sensitivity analysis, achieve sub-10$^{-2}$ prediction errors for efficient simulation approximation. Its two-stage hierarchical inference architecture constrains geometric and shear modulus parameters using stress-free state and stretching data in Stage I and enables full-parameter identification via membrane fluctuation and relaxation tests in Stage II. Applied to healthy and malaria-infected RBCs, the RBC-MsUQ framework produces statistically robust posterior distributions, revealing increased stiffness and viscosity in pathological cells. Quantitative model-experiment validation demonstrates that RBC-MsUQ effectively mitigates uncertainties through cross-platform data fusion, overcoming the critical limitations of existing computational approaches. The RBC-MsUQ framework thus provides a systematic paradigm for studying RBC properties and advancing cellular mechanics and biomedical engineering.
\end{abstract}


%Research highlights
\iffalse
\begin{highlights}
\item RBC-MsUQ integrates multi-platform data via hierarchical Bayesian inference.
\item Key RBC morpho-mechanical parameters are refined in multi-stage engineering-based manner.
\item Neural networks are utilized to develop fast surrogates for mechanical simulations.
\item Validation demonstrate RBC-MsUQ's efficacy in uncertainty quantification.
\item RBC-MsUQ is extensible to analyze morpho-mechanics in natural and bioengineered cellular systems.
\end{highlights}
\fi

%% Keywords

\end{frontmatter}

\section{Introduction}
\label{sec_1}
The morpho-mechanical characterization of biological cells constitutes a fundamental challenge at the intersection of computational mechanics and biomedical engineering, with profound implications for understanding pathophysiological mechanisms and advancing therapeutic technologies. Although traditional engineering materials exhibit well-defined constitutive relationships, biological systems such as red blood cells (RBCs) present unique complexities because of their active viscoelasticity, multiscale architecture, and dynamic environmental interactions. RBCs, as the primary oxygen transporters in circulation, exemplify nature's optimization of deformability---capable of navigating microcapillaries as narrow as \SI{3}{\micro\metre} through cyclic shape transformations. This extraordinary functionality arises from the synergistic interplay between membrane elasticity, cytoskeletal dynamics, and cytoplasmic viscosity---parameters that are critically sensitive to hematological disorders such as malaria, sickle cell anemia (SCA) and hereditary disorders~\cite{rowe2009adhesion,haldar2007erythrocyte,miller2002pathogenic,cooke2001malaria,bannister2000brief,millholland2011malaria,aingaran2012host}. Pathological alterations in these properties alter microvascular flow, promote thrombogenesis, and exacerbate organ dysfunction, underscoring the urgency of establishing robust mechano-biological characterization frameworks.

%%%%%%%%%%%%%%%%%%%%%%%%%%%%%% Fig1 Overall schematic %%%%%%%%%%%%%%%%%%%%%%%%%%%%%%
\begin{figure}[!ht]
\begin{center}
\includegraphics[width=1.00\textwidth]{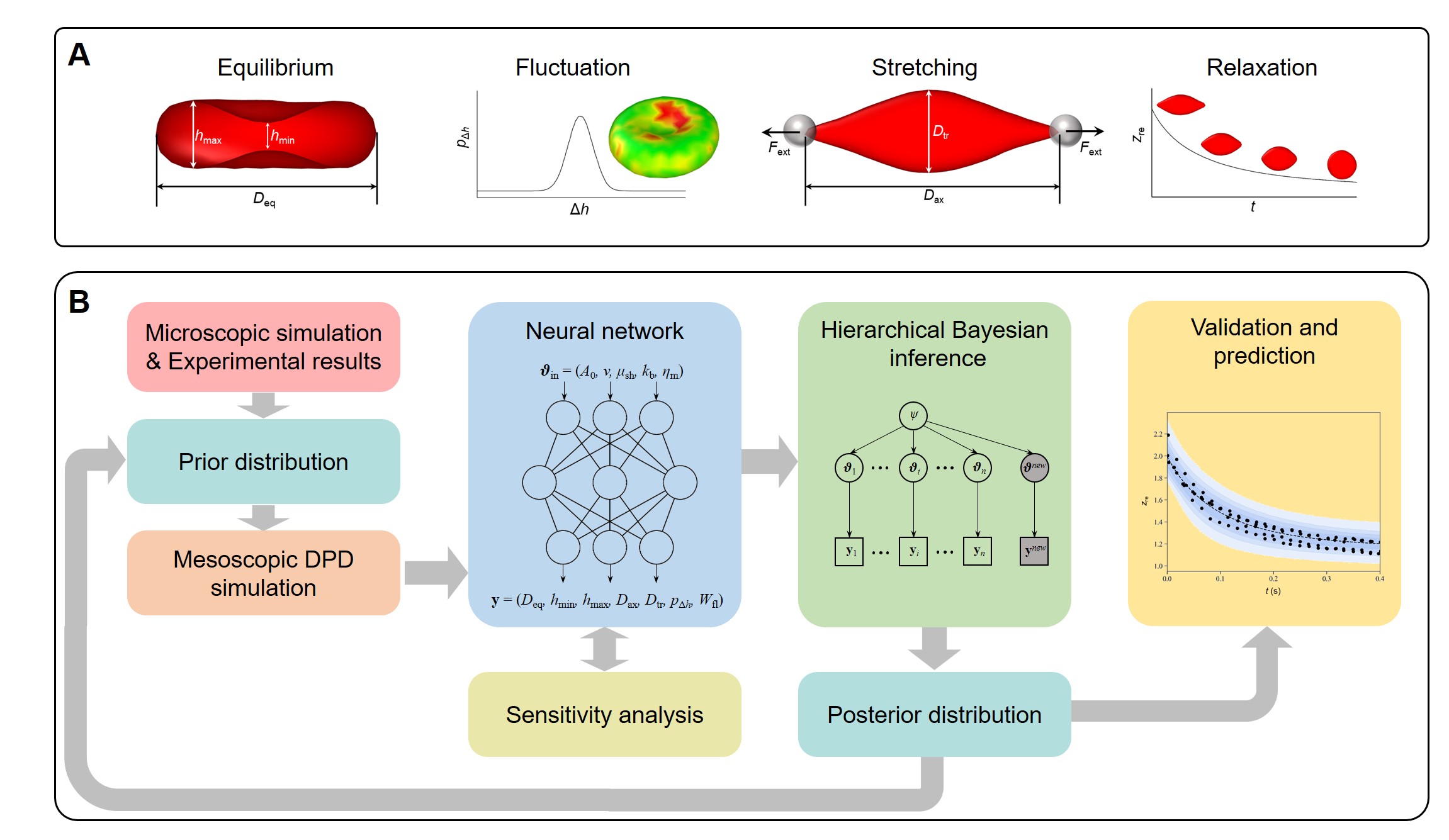}
\end{center}
\vspace{-0.25in}
\caption{\small{\bf Schematic of the RBC-MsUQ modeling framework}. (A) Overview of four representative experiments employed to inform parameter inference within our model: equilibrium shape at a stress-free state, membrane fluctuations upon substrate adhesion, deformation under applied stretching forces, and relaxation dynamics following force release. (B) Structure of the RBC-MsUQ framework. Mesoscopic DPD simulations, designed to replicate experimental scenarios, incorporate morpho-mechanical properties $\boldsymbol{\vartheta}^{\mathrm{in}}=(A_0,v,\mu_{\mathrm{sh}},k_{\mathrm{b}},\eta_{\mathrm{m}})$ derived from prior distributions established through microscopic simulations and available experimental data on both healthy and diseased RBCs. Surrogate neural network models are subsequently trained on these parameters and the resulting simulation results $\bm{\mathrm{y}}=(D_{\mathrm{eq}},h_{\mathrm{max}},h_{\mathrm{min}},D_{\mathrm{ax}},D_{\mathrm{tr}},t_\mathrm{c},W_\mathrm{fl})$. This procedural approach facilitates two-stage hierarchical Bayesian inferences, yielding posterior distributions of $\boldsymbol{\vartheta}^{\mathrm{in}}$ relevant to the structural and mechanical properties of RBCs in health and disease. The model's predictions, based on the refined parameter distributions and surrogate models, demonstrate alignment with experimental data, with further validation supported by observed RBC behavior during splenic slit traversals.}
\label{fig:1}
\end{figure}
%%%%%%%%%%%%%%%%%%%%%%%%%%%%%%%%%%%%%%%%%%%%%%%%%%%%%%%%%%%%%%%%%%%%%%%%%%%%%%%%%%%%

%%%%%%%%%%%%%%%%%%%%%%%%% experiments %%%%%%%%%%%%%%%%%%%%%%%%
Significant experimental efforts have been devoted to elucidating the morpho-mechanical characteristics of healthy and diseased RBCs through a variety of methodologies, including atomic force microscopy (AFM), optical tweezers, micropipette aspiration, and microfluidic deformability assays~\cite{depond2020methods,aikawa1996membrane,nagao2000plasmodium,li2006observations,dearnley2016reversible,sisquella2017plasmodium,perez2018multispectral}. For example, Evans et al. standardized critical measurements of RBC dimensions using an interference microscope as early as 1972~\cite{evans1972improved}. Subsequent technological advances have introduced a variety of imaging techniques---ranging from digital holographic microscopy to serial block-face scanning electron microscopy---to facilitate the three-dimensional (3D) reconstruction of healthy RBCs (hRBCs) and Plasmodium falciparum-infected RBCs (Pf-RBCs)~\cite{esposito2010quantitative,anand2012automatic,agnero2019malaria,sakaguchi2016three,LOH2021101845}. In addition, other technologies such as ektacytometry, microsphiltration, and imaging flow cytometry are being developed to investigate the morpho-mechanical properties of hRBCs and Pf-RBCs, providing insights into the structural and biomechanical changes of RBCs under parasite invasion~\cite{maier2008exported,cranston1984plasmodium,deplaine2011sensing,duez2018high,barber2018reduced,safeukui2013surface}.
%
%%%%%%%%%%%%%%%%%%%%% computational models %%%%%%%%%%%%%%%%%%%%
Along with the aforementioned experimental studies, recent advances in computational modeling and simulation enable the investigation of a wide range of morpho-mechanical problems associated with RBCs at various spatial-temporal scales. Several computational models, including continuum models and particle-based methods, have been developed to explore the behaviors of RBCs in greater depth~\cite{zhou2005finite,aingaran2012host,pivkin2008accurate,fedosov2010multiscale,fedosov2011quantifying,fedosov2011wall,ye2013stretching,chang2016md,zhang2015multiple,dearnley2016reversible}. For example, parallel advances in computational modeling, including coarse-grained molecular dynamics (CGMD), dissipative particle dynamics (DPD) and smoothed particle hydrodynamics (SPH), have enabled multiscale simulations of RBC dynamics under physiological flow conditions~\cite{wu2013simulation,hosseini2012malaria,Li_2012_BJ,imai2010modeling,Li_2014_BJ} These models successfully replicate phenomena such as malaria-induced membrane stiffening and sickle cell shape transitions, providing critical insights into disease progression~\cite{chang2016md,zhang2015multiple,dearnley2016reversible}.

%%%%%%%%%%%%%%%%%%% multi-level uncertainty %%%%%%%%%%%%%%%%%%
However, various sources of uncertainty consistently compromise the predictive accuracy of computational models. The variability in the experimental measurements, which arises from different methodologies, experimental conditions, and cell origins, compounded the challenges of accurate calibration and prediction of morpho-mechanical behavior. For example, evidence indicates that there can be systematic discrepancies of up to 100\% in the shear modulus parameter distributions of red blood cells derived from atomic force microscopy measurements compared to micropipette aspiration experiments~\cite{barns2017investigation,dulinska2006stiffness,lekka2005erythrocyte,girasole2012structure,li2012atomic,suresh2005connections,suresh2006mechanical,park2008refractive,nash1989abnormalities}. Moreover, the coupling effect of multiple parameters may lead to ill-posed inverse problems during multi-parameter calibration. For example, it is overwhelming to calibrate individual RBC properties with the experimental results of membrane fluctuation, which are sensitive to multiple factors such as cell geometry and elasticity~\cite{park2008refractive,fedosov2010multiscale}. Furthermore, random perturbations and measurement inaccuracies within the simulations can also introduce uncertainties into the computational models. To systematically address this challenge, various uncertainty quantification (UQ) methods are developed and highly valued, including Bayesian inference, sensitivity analysis, adaptive sampling, surrogate model, etc.~\cite{zhang2020basic,psaros2023uncertainty,xiu2002modeling,xiu2003modeling,winovich2019convpde}. Bayesian inference, a cutting-edge method for addressing uncertainty quantification in computational mechanics, has been successfully applied to the constitutive modeling of biomaterials and the mechanical characterization of cells, employing a probabilistic framework to model parameter uncertainties~\cite{rudoy2010bayesian,roy2012error,roy2013probabilistic,romer2022surrogate}. However, conventional single-level Bayesian frameworks encounter substantial challenges when integrating heterogeneous experimental data: the pronounced differences in data sources and measurement techniques across varying experimental modalities (such as stress-free state measurements versus dynamic stretching tests) lead to heterogeneity in the inference results. Simple single-level Bayesian inference struggles to effectively manage such heterogeneous data.

To address this challenge, a hierarchical Bayesian framework can be introduced, which constructs multi-level probabilistic models by incorporating hyperparameter distributions, thereby providing a theoretical foundation for the statistical fusion of heterogeneous data. In particular, the hierarchical surrogate modeling approach developed by the Koumoutsakos research group has successfully achieved statistical integration of multi-parameter experimental datasets, significantly enhancing the physical consistency of the posterior distribution of red blood cell parameters~\cite{amoudruz2023stress,economides2021hierarchical,wu2019hierarchical}. Nevertheless, when strong physical couplings exist between the parameters to be inverted---such as the aforementioned problems in membrane fluctuation tests and the synergistic effects between the cell membrane elastic modulus and the surface viscosity in the relaxation tests---traditional hierarchical methods still face the challenge of ensuring stable parameter identifiability. To overcome these limitations, we introduce RBC-MsUQ---a multi-stage UQ framework for RBCs that integrates hierarchical Bayesian inference with fast surrogate models for RBC mechanics, as shown in Fig.~\ref{fig:1}. Our approach decouples the inverse problem into two stages: Stage I utilizes equilibrium state analyses and stretching data from optical tweezers experiments to constrain the geometry of RBC and shear modulus distributions, while Stage II incorporates membrane fluctuation and relaxation tests to identify the full set of viscoelastic parameters. Surrogate models optimized through sensitivity analysis provide adequate approximations of output quantities with high efficiency. Predictions based on the posterior distributions of RBC morpho-mechanical properties are consistent with experimental observations, demonstrating the feasibility of parameter identification by the RBC-MsUQ framework. Bridging the gap between experimental metrology and computational mechanics, the RBC-MsUQ framework establishes a paradigm for cellular-scale mechano-biological analysis, with direct applications in therapeutic delivery system design and hematologic disorder diagnostics. In the following sections of this study, we first introduce the viscoelastic RBC model, the dynamic annealing method, and the Bayesian inference framework. We subsequently present the results of our simulations, surrogate modeling, and the multi-stage inference processes, culminating in a summary of our key findings and their implications for advancing the understanding of RBC morpho-mechanical properties in health and disease.

%%%%%%%%%%%%%%%%%%%%%%%%%%%%%%%%%%%%%%%%%%%%%%%%%%%%%%%%%%%%%%%%%%%%%%%%%%%%%%%%%%%%%%%%%%%%%%%%%%%%%%%%%%%%%%%%%%%%%%

\section{Model and Methods}
\label{sec_2}

%%%%%%%%%%%%%%%%%%%%%%% Fig2 RBC model and dynamic annealing %%%%%%%%%%%%%%%%%%%%%%%
\begin{figure}[!ht]
\begin{center}
\includegraphics[width=0.80\textwidth]{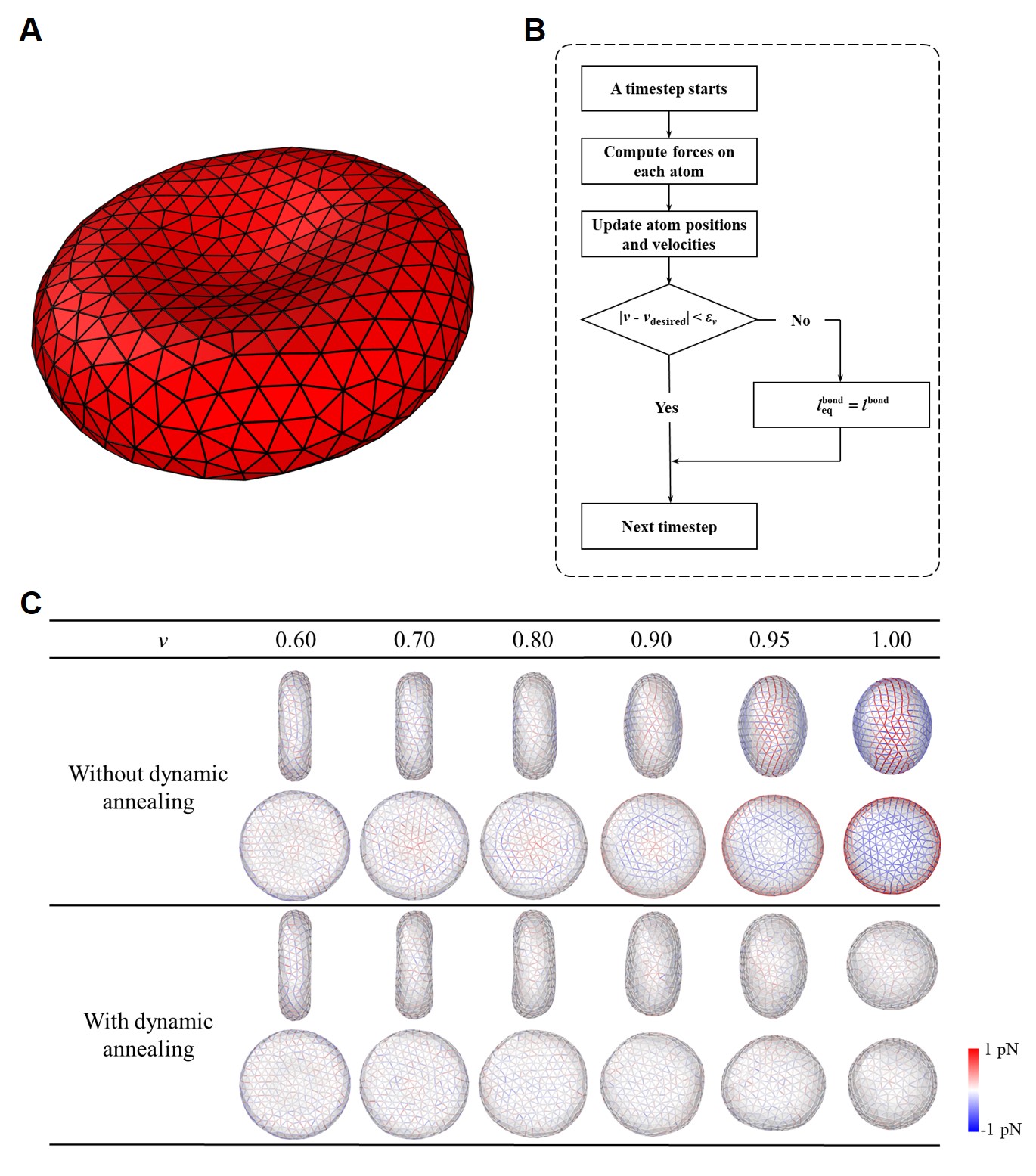}
\end{center}
\vspace{-0.25in}
\caption{\small{\bf Viscoelastic RBC model and dynamic annealing method.} (A) Schematic of the viscoelastic RBC model used in the simulations of representative mechanical experiments, featuring 500 vertices interconnected by WLC-POW bonds (represented by black lines). (B) Simplified flow chart of the dynamic annealing process designed to achieve a stress-free state. During simulation, if the difference between the current reduced volume $v$ and the desired volume $v_{\mathrm{desired}}$ exceeds the tolerance threshold ${\epsilon}_v$, the equilibrium length $l_{\mathrm{eq}}^{\mathrm{bond}}$ is adjusted to match the current bond length $l_i^{\mathrm{bond}}$. (C) Visualization of RBC shapes in two orientations, along with bond forces displayed at varying reduced volumes $v$ with and without the dynamic annealing method. Bond colors indicate the average bond force strength in equilibrium states.}
\label{fig:2}
\end{figure}
%%%%%%%%%%%%%%%%%%%%%%%%%%%%%%%%%%%%%%%%%%%%%%%%%%%%%%%%%%%%%%%%%%%%%%%%%%%%%%%%%%%%

In this section, we first present the details of our computational models, including the viscoelastic RBC model adopted in simulations, four representative experiments employed to calibrate the morpho-mechanical parameters, the dynamic annealing method used to achieve the stress-free state, and the simulation setup. Subsequently, we introduce surrogate models based on neural networks that provide approximations of the model output. Finally, we describe the method of Bayesian inference in both single-level and hierarchical manners.

\subsection{RBC model and dynamic annealing method}
\label{sec_2_1}
\subsubsection{Viscoelastic RBC model}
\label{sec_2_1_1}
To accurately characterize the morpho-mechanical properties and simulate the representative experiments of the RBCs, we employ the viscoelastic RBC model proposed by Pivkin \emph{et al.}~\cite{pivkin2008accurate,fedosov2010multiscale,peng2013lipid}, as shown in Fig.~\ref{fig:2}A. The model was developed on the basis of the dissipative particle dynamics (DPD) method, which is a coarse-grained particle-based method and has been widely utilized in the soft matter and biomechanical fields. The details of the DPD methods can be found in previous computational studies on the dynamics of RBCs and blood flow~\cite{qi2021quantitative,ma2023computational,han2023silico,li2023silico,pivkin2016biomechanics,li2018mechanics,chang2017modeling,tang2017openrbc,li2023analysis}. In this model, the RBC is assumed to be a two-dimensional triangular network composed of $N_v$ vertices connected by $N_b$ bonds. The model is governed by four interactions, including viscoelastic bonds, harmonic dihedrals, and area and volume constraints. The energy of the model is given by:
\begin{equation}
    E_\mathrm{RBC}= E_{\mathrm{e}}+E_{\mathrm{b}}+E_{\mathrm{a}}+E_{\mathrm{v}} \,
    \label{eq:E}\\
\end{equation}

%%%%%%%%%%%%%%%%%%%%% viscoelastic energy %%%%%%%%%%%%%%%%%%%%%
The elastic energy $E_{\mathrm{e,RBC}}$ consists of an attractive worm-like chain (WLC) spring and a repulsive power function (POW), which exhibits the non-linear feature of the RBC membrane:
\begin{eqnarray}
    E_{\mathrm{e}}=\sum_{i \in N_\mathrm{b}}\bigg [\frac{k_{\mathrm{B}}Tl_m(3x_i^2-2x_i^3)}{4p(1-x_i)}+\frac{k_p}{(n-1)l_i^{n-1}}\bigg],
    \label{eq:Ee}
\end{eqnarray}
where $l_i$ and $l_m$ denote the current and maximum bond length for bond $i$ respectively, with $x_i$ being the ratio of them. $k_{\mathrm{B}}T$ is the energy unit per mass and $p$ represents the persistence length. $k_p$ is the coefficient of the POW force and $n$ is the exponent. The combination of WLC and POW forces defines a nonlinear spring with a non-zero equilibrium length $l_0$. From the virial theorem, we derive the shear stress $\mu_{\mathrm{sh}}$ of the RBC model~\cite{dao2006molecularly}:
\begin{equation}
    \mu_{\mathrm{sh}} = \frac{\sqrt{3} k_\mathrm{B}T}{4p l_m x_0} \left( \frac{x_0}{2(1-x_0)^3} - \frac{1}{4(1-x_0)^2} + \frac{1}{4} \right) + \frac{\sqrt{3} k_p (n+1)}{4l_{\mathrm{eq}}^{n+1}}, 
    \label{eq:mu}\\
\end{equation}
where $x_0=l_0 / l_m$. In addition to the elastic property, the membrane viscosity is also characterized by introducing dissipative and random forces into the bond interaction, which is proposed by Espa$\tilde{\mathrm{n}}$ol~\cite{espanol1998fluid} with the following definitions:
\begin{equation}
    \mathbf{F}_{ij}^{\mathrm{D}} = -\gamma^{\mathrm{T}}\mathbf{v}_{ij}-\gamma^{\mathrm{C}}(\mathbf{v}_{ij} \times \mathbf{e}_{ij})\mathbf{e}_{ij} \ , 
    \label{equ:FD}\\
\end{equation}
\begin{equation}
    \mathbf{F}_{ij}^{\mathrm{R}}dt = \sqrt{2k_{\mathrm{B}}T} \left( \sqrt{2\gamma^{\mathrm{T}}} d \overline{\mathbf{W}_{ij}^{\mathrm{S}}} + \sqrt{3\gamma^{\mathrm{C}} - \gamma^{\mathrm{T}}} \frac{\mathrm{tr} [d\mathbf{W}_{ij}]}{3} \mathbf{1} \right) \times \mathbf{e}_{ij}\ , 
    \label{equ:FR}
\end{equation}
where $\gamma^{\mathrm{T}}$ and $\gamma^{\mathrm{C}}$ are the viscous coefficients, $\mathbf{v}_{ij}$ is the relative velocity between the vertices $i$ and $j$ with the corresponding unit vector $\mathbf{e}_{ij}$. $d \overline{\mathbf{W}_{ij}^{\mathrm{S}}}$ denotes the symmetric and unrestrained trace of $\mathrm{tr} [d\mathbf{W}_{ij}]$, which is the trace of a random matrix of independent Wiener increments. The membrane viscosity can be expressed as follows:
\begin{equation}
    \eta_m = \sqrt{3}\gamma^{\mathrm{T}}+\frac{\sqrt{3}}{4}\gamma^{\mathrm{C}}, 
    \label{eq:etam}\\
\end{equation}

%%%%%%%%%%%%%%%%%%%%%%% bending energy %%%%%%%%%%%%%%%%%%%%%%%
The bending energy is stored in the dihedrals between two adjacent triangles with a shared bond $j$, which is given in a harmonic form:
\begin{equation}
    E_{\mathrm{b}}=\sum_{j \in N_\mathrm{b}}k_{\mathrm{d}} \Big[1 - \cos(\theta_{j}-\theta_0) \Big],
    \label{Eb}\\
\end{equation}
where $k_{\mathrm{d}}$ is the bending coefficient, $\theta_{j}$ and $\theta_{0}$ are the current and the equilibrium angle of dihedral $j$ respectively. With Helfrich's macroscopic model~\cite{helfrich1973elastic}, the bending stiffness $k_{\mathrm{c}}$ is given by:
\begin{equation}
    k_{\mathrm{b}}=\frac{\sqrt{3}}{2} k_{\mathrm{d}},
    \label{kb}\\
\end{equation}

%%%%%%%%%%%%%%%%%%%%%% area and volume %%%%%%%%%%%%%%%%%%%%%%%
The area and volume constraints determine the incompressibility of the RBC membrane and intracellular fluid, and the energies are given by:
\begin{align}
  & E_{\mathrm{a}}=\frac{k_{\mathrm{a}}(A-A_0)^2}{2A_0}, \label{Ea}\\
  & E_{\mathrm{v}}=\frac{k_{\mathrm{v}}(V-V_0)^2}{2V_0}, \label{Ev}
\end{align}
where $k_{\mathrm{a}}$ and $k_{\mathrm{v}}$ are the area and volume constraint coefficients, and $A_0$ and $V_0$ are the desired area and volume, respectively. To ensure incompressibility of the membrane, the condition $k_{\mathrm{a}} \gg \mu_{\mathrm{sh}}$ should be satisfied. We introduce reduced volume $v=6V_0/(\pi(A_0/\pi)^{3/2})$, the ratio between the desired volume $V_0$ and the volume of the sphere with a surface area $A_0$, to represent the shape of the RBC.

With the viscoelastic model under these constraints, we have the capacity to simulate various mechanical behaviors of RBCs with desired properties. Among the parameters mentioned above, we select five vital ones that can systematically describe the properties of RBCs, including the desired surface area $A_0$, reduced volume $v$, shear stiffness $\mu_{\mathrm{sh}}$, bending stiffness $k_{\mathrm{b}}$ and membrane viscosity $\eta_{\mathrm{m}}$, as parameters of our RBC-MsUQ framework. We further perform a massive amount of simulations based on the prior distributions of these five variables and conduct subsequent Bayesian inferences.

%%%%%%%%%%%%%%%%%%%%%%%%%%%%%%%% Fig3 Bayesian model %%%%%%%%%%%%%%%%%%%%%%%%%%%%%%%
\begin{figure}[!tbp]
\begin{center}
\includegraphics[width=1.0\textwidth]{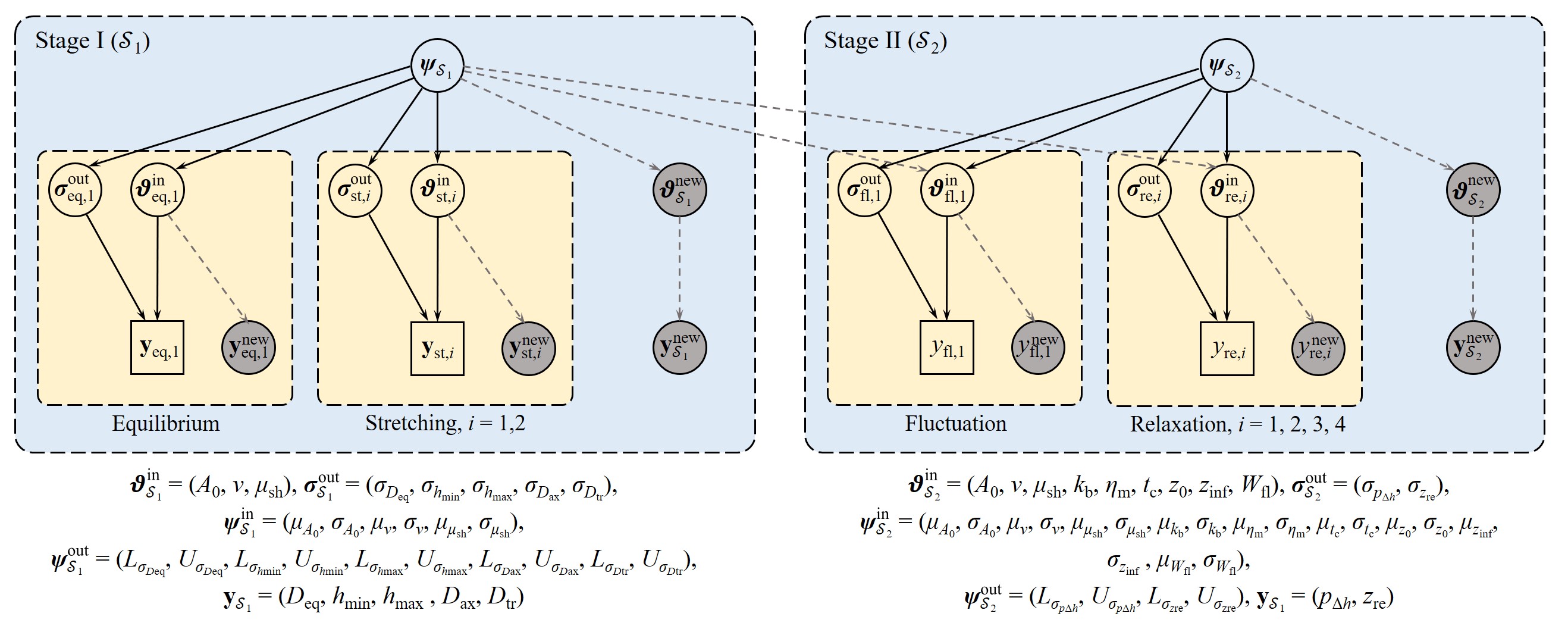}
\end{center}
\vspace{-0.25in}
\caption{\small{\bf Graphical illustration of multi-stage hierarchical Bayesian models.} The RBC-MsUQ framework consists of two hierarchical Bayesian models in a staged manner: in Stage I (left figure), the hierarchical inference is conducted based on the single-level models of equilibrium and stretching tests, with which stable distributions of a portion of the model parameters can be inferred; Subsequently in Stage II (right figure) the posterior distributions of all the model parameters are identified by the information transferred from Stage I and the fluctuation and relaxation tests. The panels shaded with light yellow and blue are the single-level and hierarchical Bayesian inference models. Here, $\boldsymbol{\vartheta}^{\mathrm{in}}$, $\boldsymbol{\sigma}^{\mathrm{out}}$, $\bm{\mathrm{y}}$, and $\bm{\mathrm{y}}^\mathrm{new}$ denote the model input parameters, SDs of the outputs, outputs, and predictions based on posterior distributions, respectively. $\boldsymbol{\psi}$ represents the hyperparameters that encode information sharing between datasets. The circles, squares, and shaded circles refer to unobserved, observed, and updated variables, respectively. Arrows illustrate the direction of forward propagation, while inference is performed in the reverse direction.}
\label{fig:3}
\end{figure}
%%%%%%%%%%%%%%%%%%%%%%%%%%%%%%%%%%%%%%%%%%%%%%%%%%%%%%%%%%%%%%%%%%%%%%%%%%%%%%%%%%%%

\subsubsection{Selected representative experiments}
\label{sec_2_1_2}
To calibrate the morpho-mechanical properties of RBCs, we need to link experimental datasets and simulation results to update the RBC-MsUQ framework. In this work, we choose four representative experiments, including measurement of the shape of the RBC in the stress-free state, membrane fluctuation when adhered to a substrate, deformation under stretching forces, and relaxation time after extension release. Evans and Fung measured the geometry of healthy RBCs with an interference microscope and obtained equilibrium diameter $D_{\mathrm{eq}}=7.82\pm0.62\mathrm{\upmu m}$, maximum thickness $h_{\mathrm{max}}=2.58\pm0.27\mathrm{\upmu m}$, and minimum thickness $h_{\mathrm{min}}=0.81\pm0.35\mathrm{\upmu m}$~\cite{evans1972improved}. For Pf-RBCs, we infer that $D_{\mathrm{eq}}=6.9\mathrm{\upmu m}$, and $h_{\mathrm{max}}=h_{\mathrm{min}}=3.2\mathrm{\upmu m}$ based on Esposito's measurement of surface area and volume~\cite{esposito2010quantitative}. Mills et al. applied extension forces on RBCs with optical tweezers and obtained axial and transverse deformations, $D_{\mathrm{ax}}$ and $D_{\mathrm{tr}}$, against different stretching forces~\cite{mills2004nonlinear}. They also measured the deformation of Pf-RBCs in three stages under stretching in the following work~\cite{suresh2005connections}. Hochmuth et al. observed the exponential time-dependent behavior of healthy RBCs upon release and introduced a characteristic time $t_{\mathrm{c}}$ to describe the relaxation process~\cite{hochmuth1979red}. The ratio between the relaxation times of hRBCs and Pf-RBCs was measured by Handayani et al. with cells passing through microfluidic channels~\cite{handayani2009high}. Park et al. extracted nanoscale cell membrane fluctuations with diffraction phase microscopy and plotted histograms of cell thickness fluctuations of hRBCs and Pf-RBCs in different stages~\cite{park2008refractive}. With these experimental datasets, we conduct corresponding simulations, construct surrogate models, and infer the distribution of morpho-mechanical properties of hRBCs and Pf-RBCs in accordance with experimental observations.

%%%%%%%%%%%%%%%%%%%%%%%%%%%%%%%%%%%%%%%%%%%%%%%%%%%%%%%%%%%%%%%%%%%%%%%%%%%
\begin{longtable}{llccc}
	\small \\
	%\centering \\
	%\arrayrulecolor{blue} \\
	%\renewcommand\arraystretch{1.5} \\
	\caption{Datasets of four representative experiments}~\label{table:Exp} \\
	%\centering \\
	%\begin{tabular}{m{80pt}<{\raggedright}  m{150pt}<{\raggedright}  m{30pt}<{\raggedright}  m{30pt}<{\raggedright}  m{40pt}<{\raggedright}}
	\cmidrule[1.5pt]{1-5}
	Cell type & Experiment & index & format & source \\  \cmidrule[1.15pt]{1-5}
	\multirow{8}{*}{hRBC} 
	& Equilibrium       & $i=1$         & $D_{\mathrm{eq}}$, $h_{\mathrm{min}}$, $h_{\mathrm{max}}$     & Evans et al.~\cite{evans1972improved}  \\
	& Stretching       & $i=1$         & $D_{\mathrm{ax}}$, $D_{\mathrm{tr}}$     & Mills et al.~\cite{mills2004nonlinear}  \\
	& Stretching       & $i=2$         & $D_{\mathrm{ax}}$, $D_{\mathrm{tr}}$     & Suresh et al.~\cite{suresh2005connections}  \\
	& Fluctuation       & $i=1$         & $p_{\Delta h}$     & Park et al.~\cite{park2008refractive}  \\
	& Relaxation       & $i=1$         & $z_{\mathrm{re}}$     & Hochmuth et al.~\cite{hochmuth1979red}  \\
	& Relaxation       & $i=2$         & $z_{\mathrm{re}}$     & Hochmuth et al.~\cite{hochmuth1979red}  \\
	& Relaxation       & $i=3$         & $z_{\mathrm{re}}$     & Hochmuth et al.~\cite{hochmuth1979red}  \\
	& Relaxation       & $i=4$         & $z_{\mathrm{re}}$     & Hochmuth et al.~\cite{hochmuth1979red}  \\
	\hline
	\multirow{4}{*}{Pf-RBC} 
	& Equilibrium       & $i=1$         & $D_{\mathrm{eq}}$, $h_{\mathrm{min}}$, $h_{\mathrm{max}}$     & Esposito et al.~\cite{esposito2010quantitative}  \\
	& Stretching       & $i=1$         & $D_{\mathrm{ax}}$, $D_{\mathrm{tr}}$     & Suresh et al.~\cite{suresh2005connections}  \\
	& Fluctuation       & $i=1$         & $p_{\Delta h}$     & Park et al.~\cite{park2008refractive}  \\
	& Relaxation       & $i=1$         & $t_{\mathrm{c}}$     & Handayani et al.~\cite{handayani2009high}  \\
	
	\cmidrule[1.5pt]{1-5}
	%\end{tabular}
\end{longtable}
%%%%%%%%%%%%%%%%%%%%%%%%%%%%%%%%%%%%%%%%%%%%%%%%%%%%%%%%%%%%%%%%%%%%%%%%%%%

\subsubsection{Dynamic annealing}
\label{sec_2_1_3}
To mimic the experimental conditions in our simulations, the stress-free state, as the basis for four representative experiments, is vital and requires high precision. In our simulations, RBCs with different reduced volumes $v$ are deformed from the initial biconcave shape with $v=0.64$. However, when $v$ is high, which means that the cell is nearly spherical, the surface stress can be relatively high and the stress-free condition is not satisfied, as shown in the upper row of Fig.~\ref{fig:2}C. The colors of the bonds on the cell surface represent the average bond force. When $v=1$, the shape of the RBC is still distinct from an ideal spherical shape with significant bond forces. This is owing to the fixed mesh conformation of the triangular network, which needs to be remeshed when the shape of the model has a remarkable variance. Hence, we develop a dynamic annealing method based on the original annealing approach proposed by Fedosov et al. to reach the stress-free state and obtain the desired shape of the RBCs~\cite{fedosov2011quantifying,pan2011predicting}. Unlike their one-time annealing when generating the initial mesh conformation, we conduct annealing dynamically during simulations, as shown in Fig.~\ref{fig:2}B. When the difference between the current reduced volume $v$ and our desired one $v_{\mathrm{desired}}$ is larger than the critical value ${\epsilon}_v$, the model is considered in non-equilibrium state, and the equilibrium length $l_{\mathrm{eq}}^{\mathrm{bond}}$ of the WLC-POW bonds will be reset to the current bond length $l_{\mathrm{eq}}^{\mathrm{bond}}$ to eliminate surface stress. With this approach, the mesh conformation will gradually adjust to fit the desired $v$, so the stress-free state can be reached. After applying the dynamic annealing method, we successfully obtain the desired cell shape, and the bond force will remain at a rather low level, as shown in the lower row of Fig.~\ref{fig:2}C. The selection of the tolerance threshold ${\epsilon}_v$ is also vital, as too high ${\epsilon}_v$ will result in incomplete annealing, while the cell with low $v$ will be too flat under too low ${\epsilon}_v$. We measure the average bond force on the cell surface with different ${\epsilon}_v$, as shown in Fig.~S1 in the Supplementary materials. Taking into account the RBC shape transition and surface stress, we adopt ${\epsilon}_v=0.002$ in our model, ensuring that the RBC is in a stress-free state during our equilibrium simulations.

\subsubsection{Simulation setting}
\label{sec_2_1_4}
In our simulations, four representative experiments are performed sequentially, as shown in Fig.~\ref{fig:1}A. Initially, the viscoelastic model of the RBC will be assigned with the desired morpho-mechanical properties, and the cell will be zoomed in to make the surface area equal to the desired value $A_0$. Then the simulations will run for a sufficient time without any external forces to ensure that the RBC reaches the equilibrium state. The diameter $D_{\mathrm{eq}}$, the maximum thickness $h_{\mathrm{max}}$ (green) and the minimum thickness $h_{\mathrm{min}}$ will be measured during this stage. The lower part of the RBC will then be fixed to simulate the state in which the RBC is adhered to a substrate, and the displacement $\Delta\mathrm{h}$ in $z$ directions away from their equilibrium positions of the vertices at the upper measured part of the RBC is recorded and histograms of membrane fluctuation are obtained. To save computational cost in the following surrogate model training and Bayesian inference section, the histograms are converted to full-width half-maximum (denoted as $W_\mathrm{fl}$) to represent the fluctuation of the cell membrane. After the measurement of the fluctuation, the RBC will be released and re-equilibrated for a short time. Subsequently, a pair of forces with the same force $F_{\mathrm{ext}}$ and opposite directions is applied on two sides of the RBC, mimicking the stretching operated by optical tweezers. The length of the RBC in the stretching direction $D_{\mathrm{ax}}$ and the width in the vertical direction $D_{\mathrm{tr}}$ are observed when the deformation is stable under extension. Finally, the load is released and the transition of the diameters of the RBC with time will be recorded to characterize the relaxation behavior. Like fluctuation, the time-dependent curve is also converted to an exponential form to save computational cost~\cite{hochmuth1979red}. Hence, relaxation behavior can be characterized by relaxation time $t_\mathrm{c}$. With the simulations that contain the four representative experimental results completed, we obtain the output quantities $\bm{\mathrm{y}}=(D_{\mathrm{eq}},h_{\mathrm{max}},h_{\mathrm{min}},D_{\mathrm{ax}},D_{\mathrm{tr}},t_\mathrm{c},W_\mathrm{fl})$ from the input parameters $\boldsymbol{\vartheta}=(A_0,v,\mu_{\mathrm{sh}},k_{\mathrm{b}},\eta_{\mathrm{m}})$, which will be used in the following surrogate model training and inferring parameter distributions based on experimental datasets.

\subsection{Surrogate neural network model}
\label{sec_2_2}
During the Bayesian inference we will introduce in the next subsection, sampling of parameters is frequently conducted, while it will cost enormous computational power if we run a simulation every time we need the model output. Hence, a surrogate model that can quickly and easily obtain output quantities from the input parameters is necessary. In this work, we construct eight surrogate models using neural networks (NNs) of the four representative experiments for hRBCs and Pf-RBCs. The fully connected NNs consist of three hidden layers, and the number of neurons varies for different experiments based on the training process. The activation and loss functions are in hyperbolic tangent and mean square error form, respectively. The neurons are optimized with an Adaptive Moment Estimation (Adam) algorithm, and we adopt a step learning rate method to quickly update the model. 10,000 simulation results are used to train each model, and the accuracy of the NNs can be found in Section~\ref{sec_3_2}.

\subsection{Bayesian inference}
\label{sec_2_3}
In this subsection, we will introduce the Bayesian inference method and the hierarchical model with which we infer the posterior distributions of the input parameters based on experimental datasets~\cite{wu2019hierarchical,economides2021hierarchical,amoudruz2023stress}. The single-level Bayesian model processing a single dataset is first presented, followed by the hierarchical one that introduces hyperparameters to deal with multiple datasets and then the hierarchical structure applied in our work. Inferences are conducted using the Korali package developed by Koumoutsakos' group~\cite{martin2022korali}.
\subsubsection{Single-level Bayesian inference}
\label{sec_2_3_1}
For a quantity of interest $\bm{\mathrm{y}}$, we assume a stochastic form which is given by:
\begin{equation}
    y_j=f(\bm{\mathrm{x}}_j,\boldsymbol{\vartheta})+\sigma\epsilon_j, \quad j=1,...,N
    \label{UQ_y}\\
\end{equation}
where $\bm{\mathrm{x}}$ is a set of input quantities, $\boldsymbol{\vartheta}$ denotes the model parameters, and $N$ represents the output dimension. All sources of errors and uncertainty of the model are assumed to be the stochastic part $\sigma\epsilon_j$ that follows a Gaussian distribution with a zero mean and a $\sigma$ standard deviation (SD). In our RBC-MsUQ framework, $\boldsymbol{\vartheta}$ is the set of vital morpho-mechanical properties we mentioned in Section~\ref{sec_2_1_1}, including the desired surface area $A_0$, reduced volume $v$, shear stiffness $\mu_{\mathrm{sh}}$, bending stiffness $k_{\mathrm{b}}$ and membrane viscosity $\eta_{\mathrm{m}}$. $\bm{\mathrm{x}}$ includes the input simulation settings, such as the stretching forces $F_{\mathrm{ext}}$, and the parameters unchanged in our RBC models such as $x_0$ in the WLC-POW bond forces, $k_a$ and $k_v$ in the area and volume constraints, etc. $\bm{\mathrm{y}}$ is the result of the four representative experiments we mentioned in Section~\ref{sec_2_1_2}, including $D_{\mathrm{eq}}$, $h_{\mathrm{max}}$, $h_{\mathrm{min}}$, $D_{\mathrm{ax}}$, $D_{\mathrm{tr}}$, $t_\mathrm{c}$, and $W_\mathrm{fl}$.

The distributions of the parameters $\boldsymbol{\vartheta}$, which describe the morpho-mechanical properties of RBCs, are of our interest, and the experimental datasets $\bm{\mathrm{d}}$ which contain the input and output data pairs can help us update the distribution according to Bayes’ theorem:
\begin{equation}   p(\boldsymbol{\vartheta}|\bm{\mathrm{d}})=\frac{p(\bm{\mathrm{d}}|\boldsymbol{\vartheta})p(\boldsymbol{\vartheta})}{p(\bm{\mathrm{d}})}
    \label{UQ_p1}\\
\end{equation}
where $p(\bm{\mathrm{d}}|\boldsymbol{\vartheta})$ is the likelihood function that represents the probability of observing experimental data $\bm{\mathrm{d}}$ with a given parameter $\boldsymbol{\vartheta}$, and $p(\bm{\mathrm{d}})$ is the evidence from the model. $p(\boldsymbol{\vartheta})$ denotes the prior probability density, and in this work the prior distributions of $\boldsymbol{\vartheta}$ are assumed to be uniform distributions, and the ranges are obtained from mesoscopic simulations and previous experiments and further adjusted according to a small-scale test, as we will discuss in the following Section~\ref{sec_3_1}. The distributions of output quantities are then expressed as follows:
\begin{equation}
    p(\bm{\mathrm{y}}^\mathrm{new}|\bm{\mathrm{d}})=\int p(\bm{\mathrm{y}}^\mathrm{new}|\boldsymbol{\vartheta})p(\boldsymbol{\vartheta}|\bm{\mathrm{d}})d\boldsymbol{\vartheta}
    \label{UQ_p2}\\
\end{equation}

A graph representation of the single-level Bayesian model is shown in the light yellow panels in Fig.~\ref{fig:3}. The unobserved, observed, and updated variables are presented by circles, squares, and shaded circles. The arrows denote the forward directions of the model and the inferences are carried out in the reverse directions.

%%%%%%%%%%%%%%%%%%%%%%%%%%%%%% Fig4 Simulation results %%%%%%%%%%%%%%%%%%%%%%%%%%%%%
\begin{figure}[!tbp]
\begin{center}
\includegraphics[width=1.0\textwidth]{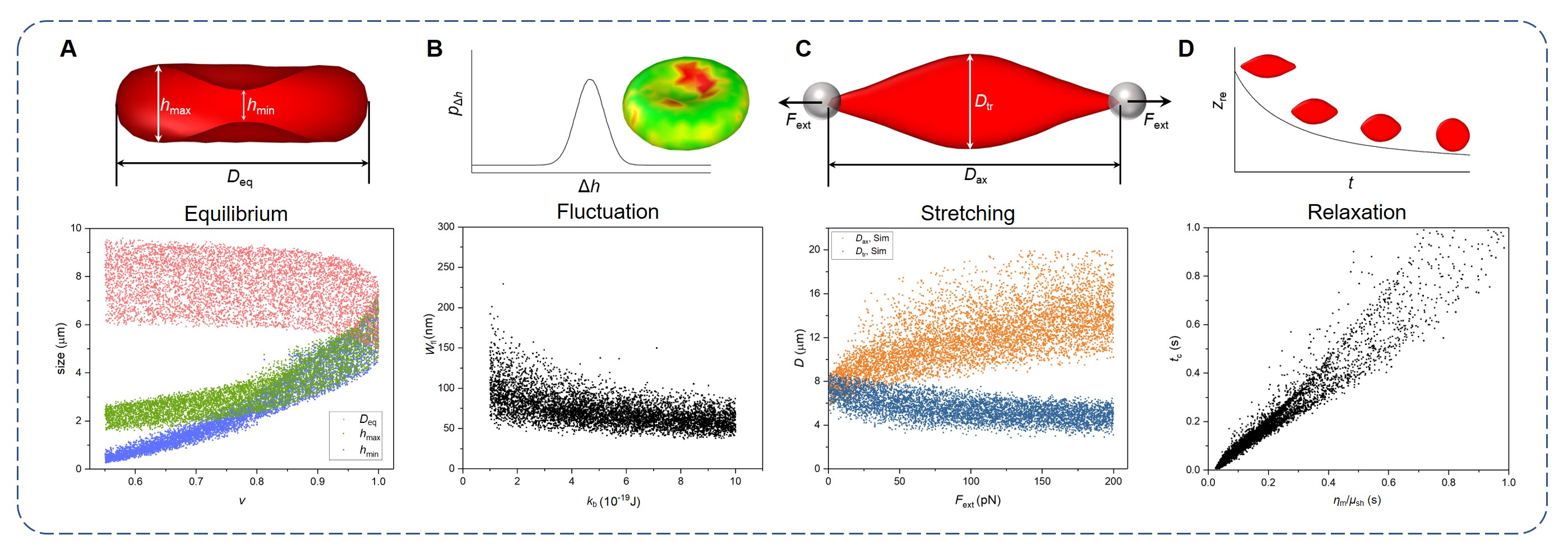}
\end{center}
\vspace{-0.25in}
\caption{\small{\bf Simulation results from four representative mechanical experiments.} (A) Equilibrium diameters $D_{\mathrm{eq}}$ (pink), maximum thickness $h_{\mathrm{max}}$ (green), and minimum thickness $h_{\mathrm{min}}$ (blue) of RBCs in a stress-free state as a function of reduced volume $v$. (B) Full-width at half-maximum ($W_\mathrm{fl}$) of membrane fluctuations as a function of bending stiffness $k_{\mathrm{b}}$ for RBCs adhered to a substrate. (C) Axial deformation $D_{\mathrm{ax}}/D_{\mathrm{eq}}$ (orange) and transverse deformation $D_{\mathrm{tr}}/D_{\mathrm{eq}}$ of RBCs subjected to external stretching forces $F_{\mathrm{ext}}$. (D) Characteristic relaxation time $t_{\mathrm{c}}$ of RBCs following stretching in relation to the ratio of membrane viscosity $\eta_{\mathrm{m}}$ to shear stiffness $\mu_{\mathrm{sh}}$. A total of 6400 simulation results based on the prior distribution of healthy RBCs are presented here, while the results for Pf-RBCs are excluded for clarity.} 
\label{fig:4}
\end{figure}
%%%%%%%%%%%%%%%%%%%%%%%%%%%%%%%%%%%%%%%%%%%%%%%%%%%%%%%%%%%%%%%%%%%%%%%%%%%%%%%%%%%%

\subsubsection{Hierarchical Bayesian inference}
\label{sec_2_3_2}
The single-level Bayesian inference approach demonstrates robust performance when applied to a single experimental dataset. However, when multiple datasets are considered, the single-level Bayesian model often fails to yield accurate results, particularly in scenarios characterized by considerable heterogeneity across experimental results. One possible strategy is to combine all datasets into a unified single-level Bayesian framework; however, prior studies have indicated that this approach may not provide reliable estimates, especially in the context of molecular simulations~\cite{zavadlav2019bayesian, wu2019hierarchical}. In contrast, hierarchical Bayesian inference provides a more effective solution~\cite{wu2018bayesian,wu2019hierarchical,economides2021hierarchical}, as it is better suited to handle the inherent variability between distinct experimental datasets, allowing for more accurate and comprehensive model predictions, as shown in the light blue panels in Fig.~\ref{fig:3}.

For every dataset, single-level Bayesian inferences are conducted separately, and then an additional level of hyperparameters $\boldsymbol{\psi}$ is introduced to connect them. For a collection of datasets $\vec{\bm{\mathrm{d}}}=\{\bm{\mathrm{d}}_1,...,\bm{\mathrm{d}}_{N_d} \}$, the posterior distribution of the model parameters $\boldsymbol{\vartheta}$ is given by:
\begin{equation}
	p\left(\boldsymbol{\vartheta}_i \mid \overrightarrow{\mathbf{d}}, \mathcal{M}_{\mathrm{HB}}\right)=\int p\left(\boldsymbol{\vartheta}_i \mid \boldsymbol{\psi}, \overrightarrow{\mathbf{d}}, \mathcal{M}_{\mathrm{HB}}\right) p\left(\boldsymbol{\psi} \mid \overrightarrow{\mathbf{d}}, \mathcal{M}_{\mathrm{HB}}\right) d \boldsymbol{\psi}.
	\label{UQ_p3}\\
\end{equation}
where $i=1,2,\ldots,Nd$. $\mathcal{M}_{\mathrm{HB}}$ denotes the hierarchical model, and in our work, two hierarchical models are constructed in two stages and are referred to as $\mathcal{S}_1$ and $\mathcal{S}_2$ respectively. Based on the dependency assumptions, the posterior distribution can be approximated using the Bayes theorem:
\begin{equation}
	p\left(\boldsymbol{\vartheta}_i \mid \overrightarrow{\mathbf{d}}, \mathcal{M}_{\mathrm{HB}}\right) \approx \frac{p\left(\mathbf{d}_i \mid \boldsymbol{\vartheta}_i, \mathcal{M}_{\mathrm{HB}}\right)}{N_s} \sum_{k=1}^{N_s} \frac{p\left(\boldsymbol{\vartheta}_i \mid \boldsymbol{\psi}^{(k)}, \mathcal{M}_{\mathrm{HB}}\right)}{p\left(\mathbf{d}_i \mid \boldsymbol{\psi}^{(k)}, \mathcal{M}_{\mathrm{HB}}\right)},
	\label{UQ_p4}\\
\end{equation}
where $\boldsymbol{\psi}^{(k)} \sim p\left(\boldsymbol{\psi} \mid \overrightarrow{\mathbf{d}}, \mathcal{M}_{\mathrm{HB}}\right)$ for $k=1,2,\ldots,N_s$, and $N_s$ represents the sample size that is quite large. The probability distribution $p\left(\boldsymbol{\psi} \mid \overrightarrow{\mathbf{d}}, \mathcal{M}_{\mathrm{HB}}\right)$ is subsequently evaluated:
\begin{equation}
	p(\boldsymbol{\psi} \mid \overrightarrow{\mathbf{d}}, \mathcal{M})=\frac{p\left(\overrightarrow{\mathbf{d}} \mid \boldsymbol{\psi}, \mathcal{M}_{\mathrm{HB}}\right) p\left(\boldsymbol{\psi} \mid \mathcal{M}_{\mathrm{HB}}\right)}{p\left(\overrightarrow{\mathbf{d}} \mid \mathcal{M}_{\mathrm{HB}}\right)}
	\label{UQ_p5}\\
\end{equation}
where $p\left(\boldsymbol{\psi} \mid \mathcal{M}_{\mathrm{HB}}\right)$ denotes the prior probability density function of $\boldsymbol{\psi}$ and $p\left(\overrightarrow{\mathbf{d}} \mid \mathcal{M}_{\mathrm{HB}}\right)$ represents the normalizing constant. The likelihood of dataset $i$ is further given by:
\begin{equation}
	p\left(\mathbf{d}_i \mid \boldsymbol{\psi}, \mathcal{M}_{\mathrm{HB}}\right)=\int p\left(\mathbf{d}_i \mid \boldsymbol{\vartheta}_i, \mathcal{M}_{\mathrm{HB}}\right) p\left(\boldsymbol{\vartheta}_i \mid \boldsymbol{\psi}, \mathcal{M}_{\mathrm{HB}}\right) d \boldsymbol{\vartheta}_i .
	\label{UQ_p6}\\
\end{equation}
Under assumption $p\left(\mathbf{d}_i \mid \boldsymbol{\vartheta}_i, \mathcal{M}_{\mathrm{HB}}\right) = p\left(\mathbf{d}_i \mid \boldsymbol{\vartheta}_i, \mathcal{M}_i\right)$ and Bayes’ theorem, Eq.~\ref{UQ_p6} is approximately formulated as:
\begin{equation}
	p\left(\mathbf{d}_i \mid \boldsymbol{\psi}, \mathcal{M}_{\mathrm{HB}}\right) \approx \frac{p\left(\mathbf{d}_i \mid \mathcal{M}_i\right)}{N_s} \sum_{k=1}^{N_s} \frac{p\left(\boldsymbol{\vartheta}_i^{(k)} \mid \boldsymbol{\psi}, \mathcal{M}_{\mathrm{HB}}\right)}{p\left(\boldsymbol{\vartheta}_i^{(k)} \mid \mathcal{M}_i\right)}
	\label{UQ_p7}\\
\end{equation}
where $\boldsymbol{\psi}_i^{(k)} \sim p\left(\boldsymbol{\vartheta} \mid \overrightarrow{\mathbf{y}}_i, \mathcal{M}_i\right)$ for $k=1,2, \ldots, N_s$. With hierarchical Bayesian inference, we are able to obtain better informed posterior distributions of $\boldsymbol{\vartheta}_i$ for each dataset, and a new general posterior $\boldsymbol{\vartheta}^{\rm{new}}$ is generated based on all datasets and can significantly reflect the experimental information. More details on hierarchical Bayesian models can be found in Ref.~\cite{wu2018bayesian,wu2019hierarchical,economides2021hierarchical}.

%%%%%%%%%%%%%%%%%%%%%%%%%%%%%%%%%%%%%%%%%%%%%%%%%%%%%%%%%%%%%%%%%%%%%%%%%%%%%%%%%%%%%%%%%%%%%%%%%%%%%%%%%%%%%%%%%%%%%%

\section{Results and discussion}
\label{sec_3}
In this section, we assume prior distributions for the model parameters based on microscopic simulations and previous experiments, which are subsequently refined through a preliminary Bayesian inference test. Subsequently, we conducted 10,000 simulations of four representative experiments for hRBCs and Pf-RBCs, of which the results were processed to train the corresponding surrogate NN models. Subsequently, we perform a sensitivity analysis of the model output with respect to the morpho-mechanical parameters of the RBCs based on these surrogate models. Next, the prior distributions of both types of RBCs are identified through multi-stage hierarchical Bayesian inferences, leveraging experimental datasets. Finally, we compare the model predictions and the experimental results, demonstrating the precision of the posterior distributions.

\subsection{Prior parameter range and simulation results}
\label{sec_3_1}
The range of input parameters has a notable impact on posterior distributions during Bayesian inference. Based on microscopic and mesoscopic simulations, as well as previous experiments~\cite{chang2016md,zhang2015multiple,dearnley2016reversible,barns2017investigation,dulinska2006stiffness,lekka2005erythrocyte,girasole2012structure,li2012atomic,suresh2005connections,suresh2006mechanical,park2008refractive,nash1989abnormalities}, we initially assumed prior ranges for the parameters of hRBCs and Pf-RBCs. To validate these assumptions and minimize computational cost, we conduct a small-scale set of simulations using the assumed prior distributions. We subsequently train surrogate neural networks, infer the parameters based on experimental results using hierarchical Bayesian inference, and obtain posterior distributions. Through analysis of the posterior distributions, we identify several inappropriate settings: 1) the upper bound of the surface area of hRBCs is too low, prompting a revision of the range from 70-\SI{160}{\square\micro\metre} to 70-\SI{185}{\square\micro\metre}; 2) the posterior distribution of the shear modulus of Pf-RBCs concentrates mainly near the lower end of the range, leading to a revision of the range from 40-\SI{140}{\micro\N/\metre} to 1-\SI{100}{\micro\N/\metre}; 3) the distribution of membrane viscosity for Pf-RBCs shows little presence over \SI{40}{\Pa\cdot\s\cdot\micro\metre}, prompting an update of the range from 4-\SI{80}{\Pa\cdot\s\cdot\micro\metre} to 5-\SI{40}{\Pa\cdot\s\cdot\micro\metre}. Following these adjustments, we perform massive-scale simulations based on the validated prior distributions, and the results are shown in Fig.~\ref{fig:4}. Each point represents a simulation result, and 6,400 simulations are performed for each mechanical experiment.

\subsection{Surrogate model and sensitivity analysis}
\label{sec_3_2}
\begin{figure}[!ht]
\begin{center}
\includegraphics[width=1.0\textwidth]{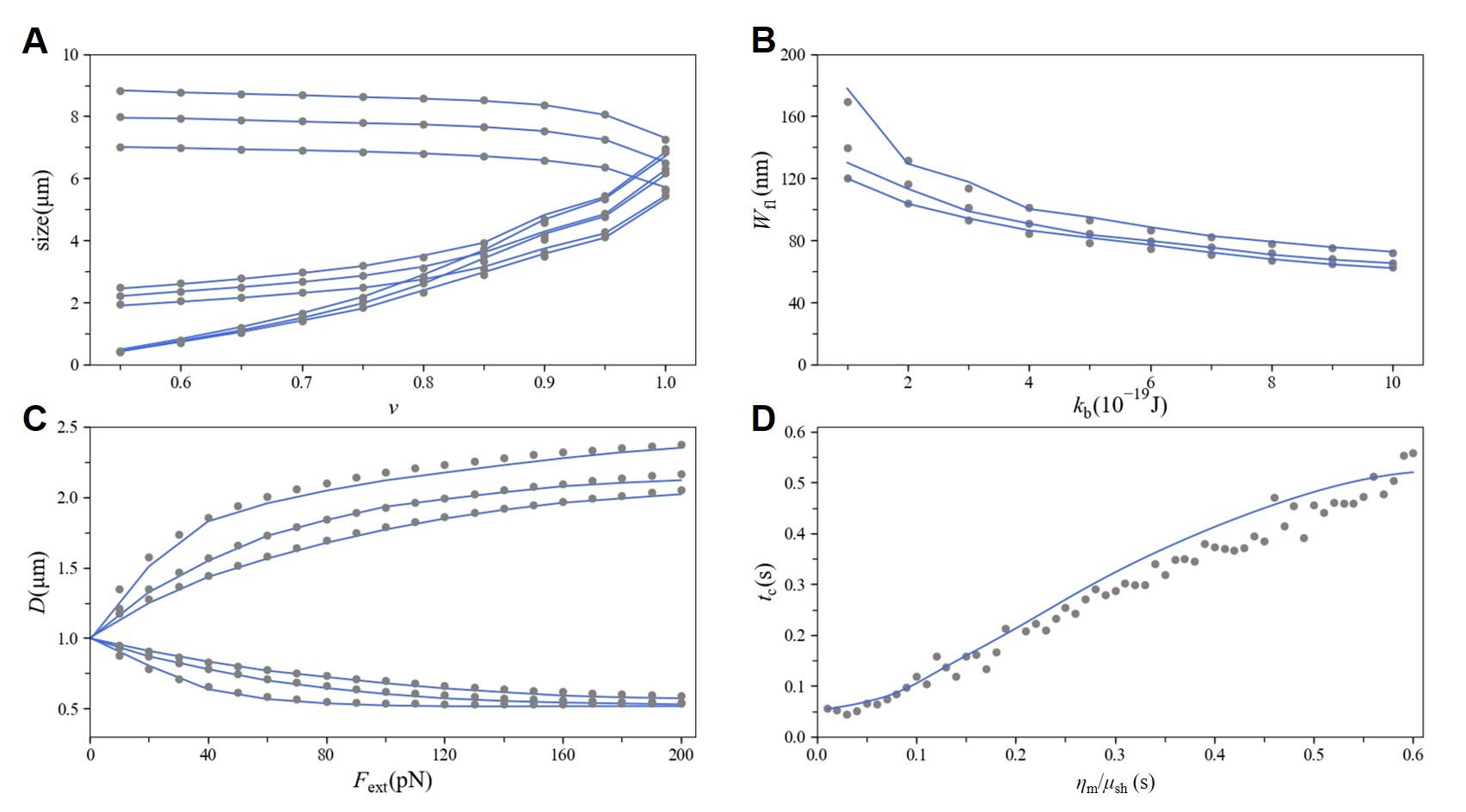}
\end{center}
\vspace{-0.25in}
\caption{\small{\bf Comparison between the neural network predictions and computational simulation results from four representative experiments.} (A) The shapes of RBC in stress-free states as a function of reduced volume $v$. (B) Membrane fluctuations as a function of bending modulus $k_\mathrm{b}$. (C) RBC deformation under stretching against different external stretching forces $F_\mathrm{ext}$. (D) Characteristic relaxation time in relation to the ratio $\eta_\mathrm{m}/\mu_{\mathrm{sh}}$ of membrane viscosity to shear modulus. The neural network (NN) predictions are depicted by blue lines, while the corresponding simulation results are presented by gray points. In panels (A-C), results corresponding to three distinct shear moduli $\mu_{\mathrm{sh}}$ are presented. For each experiment, apart from the varying parameters, all other input variables remain consistent for both the NNs and the simulations.} 
\label{fig:5}
\end{figure}

\begin{figure}[!ht]
\begin{center}
\includegraphics[width=1.0\textwidth]{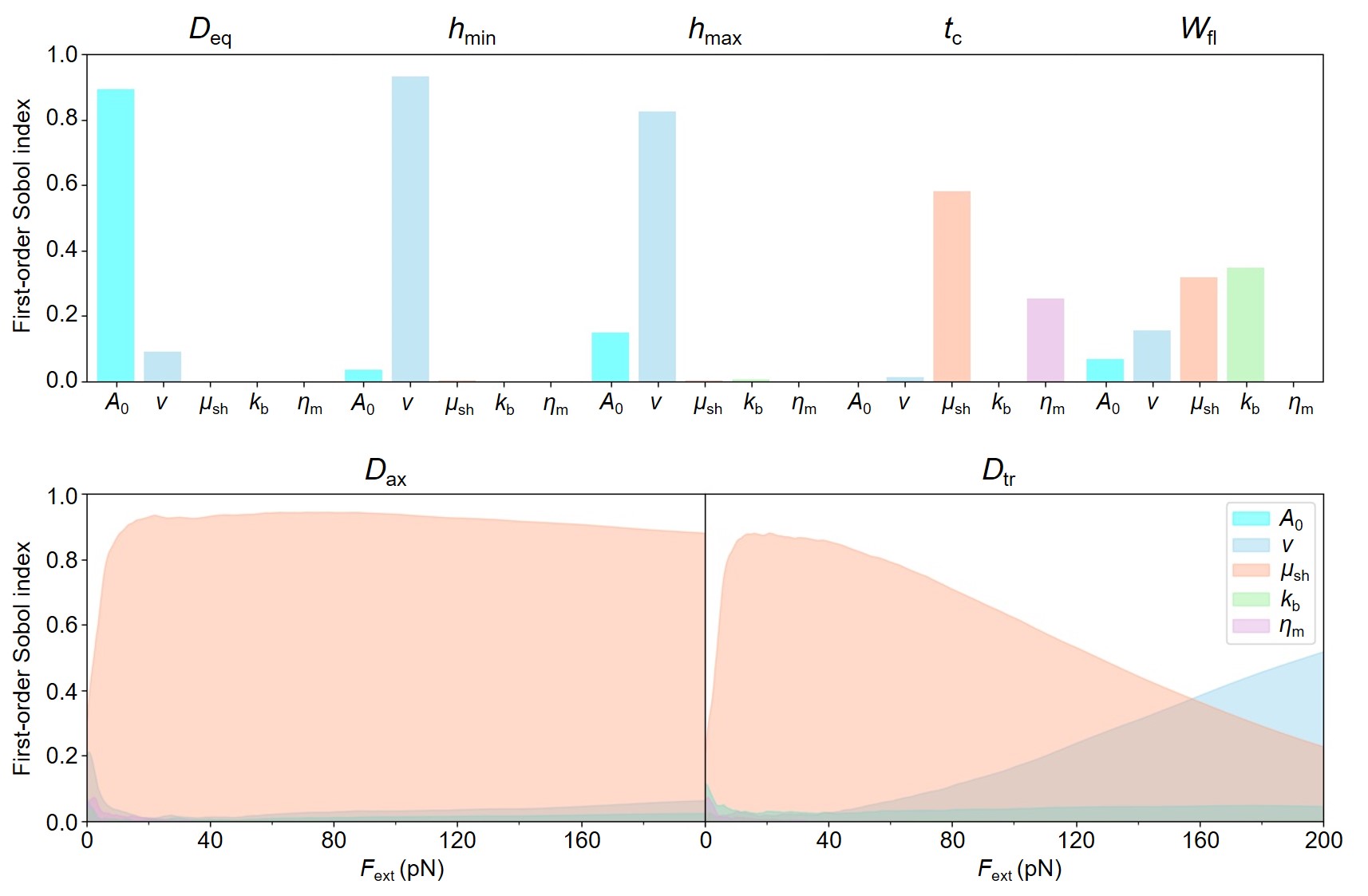}
\end{center}
\vspace{-0.25in}
\caption{\small{\bf Sensitivity analysis of model output with respect to input parameters.} The first-order Sobol index quantifies the sensitivity of the model outputs to individual input parameters, including surface area $A_0$ (cyan), reduced volume $v$ (light blue), shear stiffness $\mu_{\mathrm{sh}}$ (pink), bending stiffness $k_{\mathrm{b}}$ (light green), and membrane viscosity $\eta_{\mathrm{m}}$ (purple).} 
\label{fig:6}
\end{figure}

Upon sufficient simulation results have been obtained from four representative mechanical experiments, we construct surrogate models using neural networks. These models are trained separately on the results of the four representative experiments for hRBCs and Pf-RBCs, maintaining an error level of $10^{-2}$. The deviations between the predictions of the surrogate model (represented by the blue lines in Fig.~\ref{fig:5}) and the simulation results (depicted by the gray points) are negligible, indicating that neural networks can serve as efficient surrogate models. Subsequently, we performed a sensitivity analysis using SALib based on the surrogate model to investigate the influences of input parameters on experimental results~\cite{herman2017salib}. The first-order Sobol indices, which quantify the independent effect of each input parameter, are presented in Fig.~\ref{fig:6}. For the results of the equilibrium model $D_{\mathrm{eq}}$, $h_{\mathrm{max}}$, and $h_{\mathrm{min}}$, only the surface area $A_0$ and the reduced volume $v$ exhibit a significant impact, demonstrating that the size and shape of RBC in a stress-free state are determined primarily by surface area and volume. In the stretching test, axial deformation $D_{\mathrm{ax}}/D_{\mathrm{eq}}$ is mainly dependent on the shear modulus $\mu_{\mathrm{sh}}$, while transverse deformation is also influenced by reduced volume $v$, especially under high extension forces. The characteristic relaxation time $t_{\mathrm{c}}$ is sensitive to shear modulus and membrane viscosity, according to previous simulations and experimental observations~\cite{hochmuth1979red,ye2013stretching}. The membrane fluctuation (represented here by $W_\mathrm{fl}$) is influenced by all parameters except the membrane viscosity, which is reasonable because the fluctuation is affected not only by the in-plane and bending elasticity of the cell membrane but also by the size and shape of the RBC. Based on the results of the sensitivity analysis, irrelevant input parameters can be excluded when training surrogate models and conducting subsequent single-level Bayesian inferences.

%%%%%%%%%%%%%%%%%%% Fig5 NN prediction and simulation comparison %%%%%%%%%%%%%%%%%%%
%%%%%%%%%%%%%%%%%%%%%%%%%%%%%%%%%%%%%%%%%%%%%%%%%%%%%%%%%%%%%%%%%%%%%%%%%%%%%%%%%%%%

%%%%%%%%%%%%%%%%%%%%%%%%%%%%% Fig6 Sensitivity analysis %%%%%%%%%%%%%%%%%%%%%%%%%%%%
%%%%%%%%%%%%%%%%%%%%%%%%%%%%%%%%%%%%%%%%%%%%%%%%%%%%%%%%%%%%%%%%%%%%%%%%%%%%%%%%%%%%

\subsection{Multi-stage hierarchical Bayesian inference}
\label{sec_3_3}
After obtaining the fast neural network surrogate model, we perform Bayesian inference by incorporating experimental datasets to obtain posterior distributions of the morpho-mechanical parameters of RBCs. Given that multiple datasets from different types of experiments are available, we adopt a hierarchical Bayesian model for uncertainty quantification. First, single-level Bayesian inferences are conducted for every set of experimental data to obtain multiple distributions for the input morpho-mechanical parameters $\boldsymbol{\vartheta}^{\mathrm{in}}$. Then, hyperparameters $\boldsymbol{\psi}$ that govern the prior distributions of $\boldsymbol{\vartheta}^{\mathrm{in}}$ are constructed, ultimately leading to the posterior distribution of the parameters. However, during attempts to construct the model, we find that obtaining stable single-level Bayesian inference results for fluctuation and relaxation tests is challenging. This results from the fact that for these two tests, it is difficult to determine the distribution of each input parameter from the experimental data, as the output results are often influenced by the coupling of multiple input parameters. For example, in the case of the relaxation test, the output characteristic relaxation time $t_{\mathrm{c}}$ is close to the ratio between $\eta_{\mathrm{m}}$ and $\mu_{\mathrm{sh}}$. Therefore, when inferring the distributions of $\eta_{\mathrm{m}}$ and $\mu_{\mathrm{sh}}$ based on $t_{\mathrm{c}}$ measured from the experiment, the ratio $\eta_{\mathrm{m}}/\mu_{\mathrm{sh}}$tends to remain stable, but the distributions of the individual parameters vary significantly in repeated tests. This leads to considerable uncertainty when these results are further combined with other experimental data for hierarchical Bayesian inference.

Therefore, in this work, we design a multi-stage hierarchical Bayesian inference model, as shown in Fig.~\ref{fig:3}. We divide the four representative experimental datasets from the equilibrium and stretching tests, in which stable posterior distributions can be obtained using single-level Bayesian inference; the other group consists of the fluctuation and relaxation tests, which fail to infer stable input parameter distributions. We first perform hierarchical Bayesian inference based on the results of the equilibrium and stretching tests, obtaining the posterior distributions of RBC surface area $A_0$, reduced volume $v$, and the shear modulus $\mu_{\mathrm{sh}}$, and this process is referred to as Stage I. Subsequently, we combine the distribution information of Stage I and the experimental datasets from the fluctuation and relaxation tests to perform inference in Stage II, further identify the distributions of bending stiffness $k_{\mathrm{b}}$ and membrane viscosity $\eta_{\mathrm{m}}$. With this multi-stage strategy, stable distributions based on heterogeneous experimental information can be inferred to fully quantify the uncertainties in the computational models. In the following part of this section, we will introduce the details of the RBC-MsUQ framework and present the posterior distributions of model parameters and corresponding predictions in different stages of the model. Finally, we will perform the slit passage test based on the posterior distributions to further verify the accuracy and advantages of the RBC-MsUQ framework.

\subsubsection{Stage I: equilibrium and stretching tests}
\label{sec_3_3_1}
In Stage I, we conduct hierarchical Bayesian inference based on experimental datasets including the equilibrium shape of RBCs in stress-free state and the deformation under stretching forces, to obtain posterior distributions of surface area $A_0$, reduced volume $v$, and shear stiffness $\mu_{\mathrm{sh}}$. Initially, we start with equilibrium tests, in which the diameter $D_{\mathrm{eq}}$, minimum thickness $h_{\mathrm{min}}$, and maximum thickness $h_{\mathrm{max}}$ of hRBCs and Pf-RBCs are measured with the application of the dynamic annealing method. Under the assumption of Eq.~\ref{UQ_y}, the statistical model for the equilibrium test is:
\begin{align}
    & y_{D_{\mathrm{eq}},1} = G_{D_{\mathrm{eq}}}(\boldsymbol{\vartheta}^{\mathrm{in}}_{\mathrm{eq}}) + \sigma_{D_{\mathrm{eq}},1} \varepsilon_{D_{\mathrm{eq}},1}\ , \label{eq:H1-eq1}\\
    & y_{h_{\mathrm{min}},1} = G_{h_{\mathrm{min}}}(\boldsymbol{\vartheta}^{\mathrm{in}}_{\mathrm{eq}}) + \sigma_{h_{\mathrm{min}},1} \varepsilon_{h_{\mathrm{min}},1}\ , \label{eq:H1-eq2}\\
    & y_{h_{\mathrm{max}},1} = G_{h_{\mathrm{max}}}(\boldsymbol{\vartheta}^{\mathrm{in}}_{\mathrm{eq}}) + \sigma_{h_{\mathrm{max}},1} \varepsilon_{h_{\mathrm{max}},1}\ , \label{eq:H1-eq3}
\end{align} 
where $G_{D_{\mathrm{eq}}}(\boldsymbol{\vartheta}^{\mathrm{in}}_{\mathrm{eq}})$, $G_{h_{\mathrm{min}}}(\boldsymbol{\vartheta}^{\mathrm{in}}_{\mathrm{eq}})$, and $G_{h_{\mathrm{max}}}(\boldsymbol{\vartheta}^{\mathrm{in}}_{\mathrm{eq}})$ represent the results of the surrogate model for the equilibrium test when the input is $\boldsymbol{\vartheta}^{\mathrm{in}}_{\mathrm{eq}}$. Based on the results of the sensitivity analysis in Fig.~\ref{fig:6}, the equilibrium shape depends on the surface area and reduced volume, leading to $\boldsymbol{\vartheta}^{\mathrm{in}}_{\mathrm{eq}}=(A_0,v)$. $\sigma_{D_{\mathrm{eq}},1}$, $\sigma_{h_{\mathrm{min}},1}$, and $\sigma_{h_{\mathrm{max}},1}$ denote the standard deviations of the outputs, which is also calibrated along with $\boldsymbol{\vartheta}^{\mathrm{in}}_{\mathrm{eq}}$ in the subsequent Bayesian inference. $\varepsilon_{D_{\mathrm{eq}},1}$, $\varepsilon_{h_{\mathrm{min}},1}$, and $\varepsilon_{h_{\mathrm{max}},1}$ are independent random variables satisfying the Gaussian distribution $\mathcal{N}(0,1)$. The subscript $1$ means that there is only one experiment dataset for the equilibrium test for hRBCs and Pf-RBCs, respectively, as shown in Table~S1.

%%%%%%%%%%%%%%%%%%%%%%% Fig7 H1 Model Thetanew Prediction hRBC %%%%%%%%%%%%%%%%%%%%%
\begin{figure}[!ht]
\begin{center}
\includegraphics[width=1.0\textwidth]{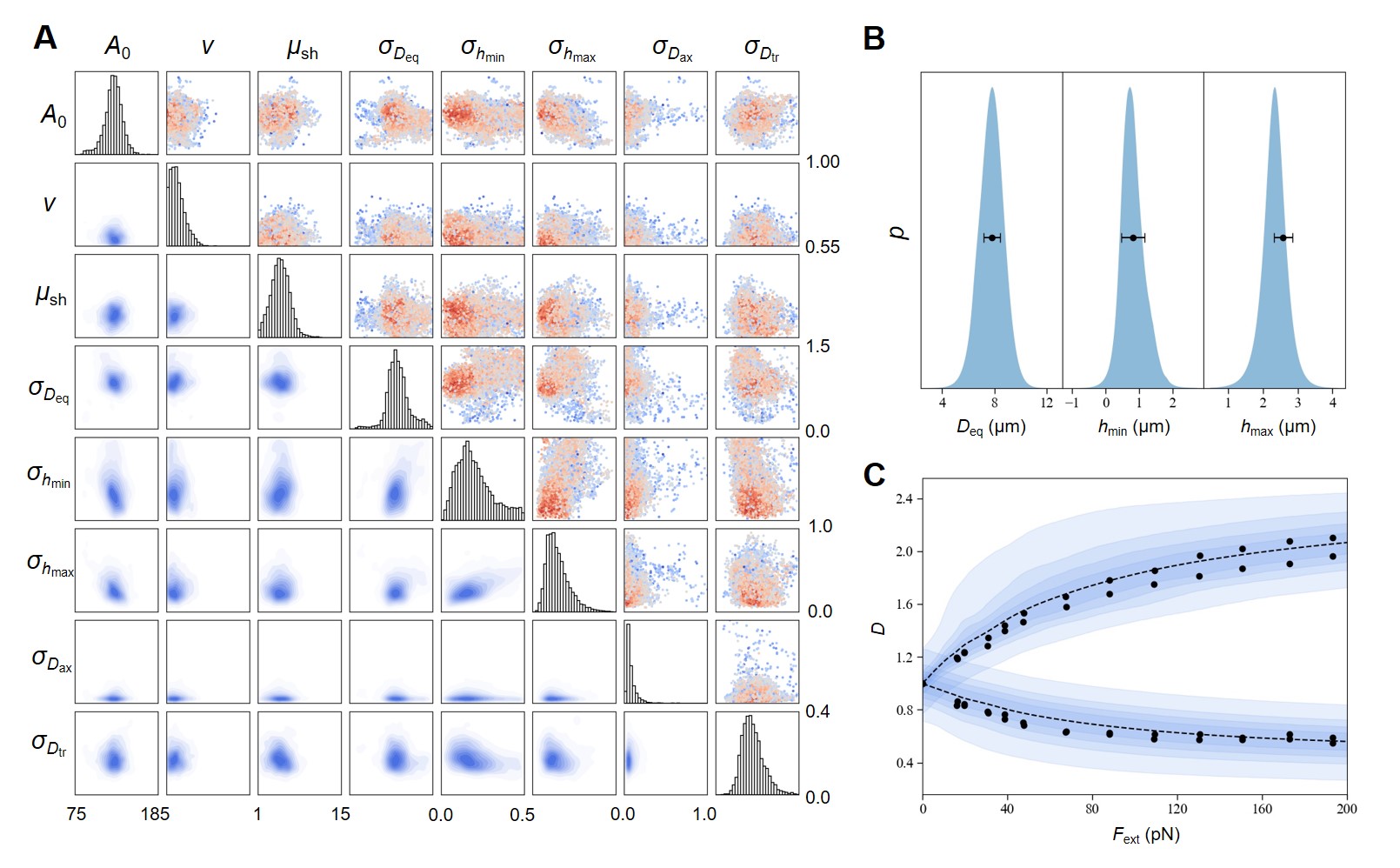}
\end{center}
\vspace{-0.25in}
\caption{\small{\bf Posterior distributions of morpho-mechanical properties and predictions in Stage I for hRBCs.} (A) The posterior distributions of the model parameters $\boldsymbol{\vartheta}_{\mathcal{S}1}$. Diagonal: marginal distributions of the parameters. Above diagonal: samples of the posterior distribution, where the color indicates the corresponding log-likelihood values (red represents high, blue represents low). Below diagonal: 2D projections of the posterior distribution colored by probability densities obtained with kernel density estimation, where darker blue indicates higher density. The ranges of parameters in the odd column are displayed at the bottom of the figure, and those in the even row are given on the right side. The means, medians, standard deviations, and maximum likelihood evaluations are summarized in Table~S1. (B-C) Predictions derived from the posterior distributions of (B) equilibrium and (C) stretching tests. The solid lines and dots denote the experimental data, and the dashed lines and shaded regions represent the mean predictions and 50{\%}, 90{\%}, 95{\%}, and 99{\%} credible intervals, respectively.} 
\label{fig:7}
\end{figure}
%%%%%%%%%%%%%%%%%%%%%%%%%%%%%%%%%%%%%%%%%%%%%%%%%%%%%%%%%%%%%%%%%%%%%%%%%%%%%%%%%%%%

For prior distributions of RBC parameters, uniform distributions are used throughout this work. The initial prior distributions are chosen as $A_0 \in \mathcal{N}(75,160)$\SI{}{\micro\square\metre} and $v \in \mathcal{N}(0.55,1.00)$ based on microscopic simulations and previous experiments. However, small-scale tests revealed that the surface area $A_0$ of healthy cells is partially distributed in the range of \SI{160}{\micro\square\metre}, and thus the previous distribution of $A_0$ for healthy erythrocytes was modified to $\mathcal{N}(75,160)$\SI{}{\micro\square\metre}. For hRBCs, the prior distributions for the standard deviations are taken as $\sigma_{D_{\mathrm{eq}}} \in \mathcal{N}(0.0,1.5)$, $\sigma_{h_{\mathrm{min}}} \in \mathcal{N}(0.0,0.5)$, and $\sigma_{h_{\mathrm{max}}} \in \mathcal{N}(0.0,1.0)$. For Pf-RBCs, the prior distributions are set as $\sigma_{D_{\mathrm{eq}}} \in \mathcal{N}(0.0,2.0)$, $\sigma_{h_{\mathrm{min}}} \in \mathcal{N}(0.0,1.0)$, and $\sigma_{h_{\mathrm{max}}} \in \mathcal{N}(0.0,1.0)$.

The results of the single-level Bayesian inference for the equilibrium shape of the RBCs in the stress-free state are presented in Fig.~S2. In the upper panel of the figure, the marginal distributions (diagonal), posterior distribution samples (above diagonal) colored by log likelihood, and 2D projections (below diagonal) colored by probability density estimated via a kernel density method are shown. For both hRBCs and Pf-RBCs, the surface area $A_0$ and the reduced volume $v$ are effectively identified from the previous uniform distributions. The mean, median, standard deviation (SD) and maximum likelihood (ML) estimates of the posterior distributions are listed in Table~S1. Given the uniform prior distribution, the maximum a posteriori (MAP) estimates coincide with the ML values, so the MAP estimates are not discussed in the following part of this work. For hRBCs, the mean values of identified $A_0$ and $v$ are \SI{134.26}{\micro\square\metre} and 0.61, respectively, which fall within the ranges observed in previous experiments. For Pf-RBCs, the posterior distributions of $A_0$ and $v$ exhibit unimodal peaks around \SI{104}{\micro\square\metre} and 0.85, respectively, indicating that Pf-RBCs lose surface areas during parasitization and assume a more spherical configuration compared to hRBCs. The lower panel of Fig.~S2 shows the predictions of the geometry of the RBC (blue area) derived from the posterior distributions, along with the experimental data (black points), demonstrating strong concordance. Due to the incomplete identification of $\boldsymbol{\sigma}^{\mathrm{out}}_{\mathrm{eq}}$ in the single-level model, the standard deviations of the output predictions around the mean values are large, leading to highly peaked distributions. These issues will be further addressed in the hierarchical inference analysis.

% stretching statistic model
The statistical model for the stretching test is as follows:
\begin{align}
    & y_{D_{\mathrm{ax}},i,j} = G_{D_{\mathrm{ax}}}(F_{\mathrm{ext},j},\boldsymbol{\vartheta}^{\mathrm{in}}_{\mathrm{st}}) + \sigma_{D_{\mathrm{ax}},i} \varepsilon_{D_{\mathrm{ax}},i,j}\ , \label{eq:RBC-MsUQ-H1-st1}\\
    & y_{D_{\mathrm{tr}},i,j} = G_{D_{\mathrm{tr}}}(F_{\mathrm{ext},j},\boldsymbol{\vartheta}^{\mathrm{in}}_{\mathrm{st}}) + \sigma_{D_{\mathrm{tr}},i} \varepsilon_{D_{\mathrm{tr}},i,j}\ , \label{eq:RBC-MsUQ-H1-st2}
\end{align}
where $G_{D_{\mathrm{ax}}}(F_{\mathrm{ax},j},\boldsymbol{\vartheta}^{\mathrm{in}}_{\mathrm{st}})$ and $G_{D_{\mathrm{tr}}}(F_{\mathrm{ax},j},\boldsymbol{\vartheta}^{\mathrm{in}}_{\mathrm{st}})$, $j=1,2,\ldots,n$ represent the results of the surrogate model for the axial deformation $D_{\mathrm{ax}}$ and the transverse deformation $D_{\mathrm{tr}}$ under the input parameters $\boldsymbol{\vartheta}^{\mathrm{in}}_{\mathrm{st}}$ and the stretching force $F_{\mathrm{ext},j}$. The subscript $i$ denotes the index of the experimental datasets, and $j$ refers to different stretching forces in the datasets, $n$ being the number of data points in a dataset. Sensitivity analysis indicates that the deformation of the RBC under stretching is dependent on the reduced volume and shear modulus, so $\boldsymbol{\vartheta}^{\mathrm{in}}_{\mathrm{st}}=(v,\mu_{\mathrm{sh}})$. In the stretching test, the prior distribution for the reduced volume remains $v \in \mathcal{N}(0.55,1.00)$, while the shear moduli for hRBCs and Pf-RBCs are set to $\mathcal{N}(1.0,15.0)$\SI{}{\micro\newton\per\metre} and $\mathcal{N}(1.0,100.0)$\SI{}{\micro\newton\per\metre} according to previous microscopic simulations and experimental observations. For healthy cells, the prior distributions for the standard errors are $\sigma_{D_{\mathrm{ax}}} \in \mathcal{N}(0.0,0.5),\sigma_{D_{\mathrm{tr}}} \in \mathcal{N}(0.0,0.2)$, while since the deformation of Pf-RBCs under stretching is smaller than hRBCs, the prior $\sigma_{D_{\mathrm{ax}}}$ and $\sigma_{D_{\mathrm{tr}}}$ for Pf-RBCs are set to $\mathcal{N}(0.0,0.05)$ and $\mathcal{N}(0.0,0.02)$.
% stretching single results
The results of the single-level Bayesian inference for the stretching tests are shown in Fig.~S3. The first two columns on the left present the results for hRBCs, derived from two sets of stretching experimental data, while the rightmost column presents the results for Pf-RBCs. The posterior distributions of reduced volume $v$, shear modulus $\mu_{\mathrm{sh}}$, and the standard deviation of the output are all well concentrated around narrow peaks.

% 1) bias between mean and ML of v 2) bias of v between two sets 3) bias of v between st and eq
For the first set of experimental data on stretching, the mean of the reduced volume $v$ is approximately 0.66, which apparently differs from the median values (0.61) and the ML values (0.59). Fig.~S3A reveals that the marginal distributions of $v$ exhibit a bimodal form with two peaks at 0.59 and 0.76. This phenomenon is similarly observed in the inference results based on the second stretching dataset for hRBCs. Analysis of the stretching simulation suggests that this variation arises from the uncertainty in the bias between the computational model and the experimental data. If we remove $D_{\mathrm{tr}}$ and only use $D_{\mathrm{ax}}$ as the model output, the inferred $v$ will be concentrated in the lower range of 0.55-0.6. In contrast, the output of $\boldsymbol{\vartheta}^{\mathrm{in}}_{\mathrm{st}}=(D_{\mathrm{tr}})$ will lead to a unimodal distribution of $v$ in the range of 0.75-0.8, especially at low axial stretching forces. This indicates a mismatch between the computational model and the experimental results, likely due to uncertainties introduced by the model itself or potential errors in the experimental datasets. Compared to the Bayesian inference results from the equilibrium test, the posterior distributions of $v$ obtained from the stretching tests differ significantly, further highlighting the heterogeneity between the different experimental results.

Moreover, the posterior distribution of the shear modulus $\mu_{\mathrm{sh}}$ for the first set of experimental data exhibits a narrow peak near \SI{5.5}{\micro\newton\per\metre}, whereas the results for the second dataset are more concentrated near \SI{3.6}{\micro\newton\per\metre}, revealing a noticeable difference. This suggests that there are discrepancies even between the results from different experimental groups of the same type, contributing to the system's uncertainty. For Pf-RBCs, the posterior distributions of $v$ and $\mu_{\mathrm{sh}}$ are characterized by narrow peaks around 0.92 and \SI{39.0}{\micro\newton\per\metre}, respectively. Consistent with the conclusions of the equilibrium tests, the stretching tests reveal that the reduced volume of Pf-RBCs is significantly larger than that of hRBCs, and the shear modulus is substantially higher, indicating that they are closer to globularity and much less deformable. The predictions of the stretching deformation of Pf-RBCs align well with the experimental values, showing excellent agreement between the model and experimental results. However, the posterior distribution of $v$ for Pf-RBCs still shows some deviation from the co-value of approximately 0.85, indicating the presence of heterogeneity between different types of experimental results.

%%%%%%%%%%%%%%%%%%%%% Fig8 H1 Model Thetanew Prediction Pf-RBC %%%%%%%%%%%%%%%%%%%%%
\begin{figure}[!ht]
\begin{center}
\includegraphics[width=1.0\textwidth]{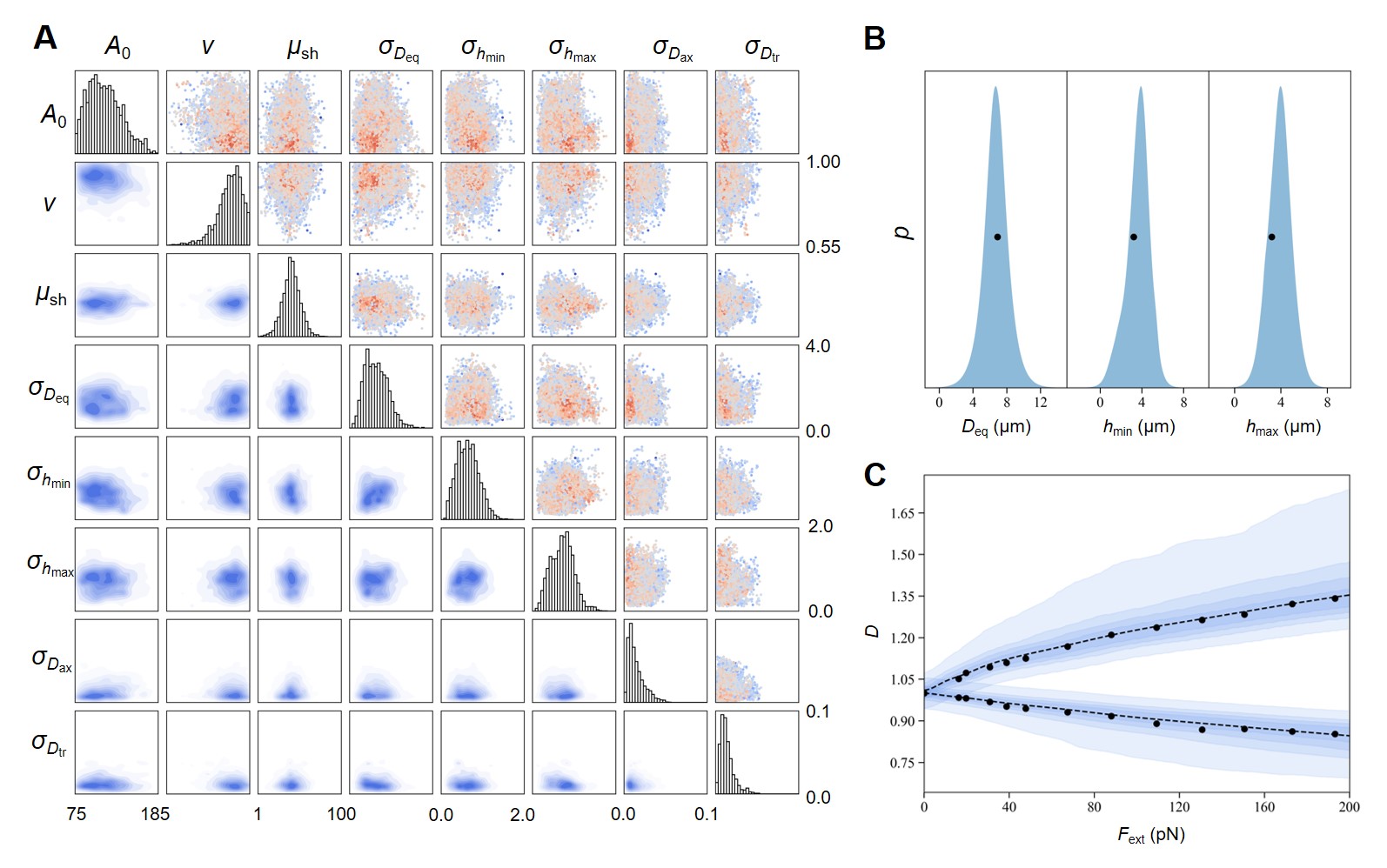}
\end{center}
\vspace{-0.25in}
\caption{\small{\bf Posterior distributions of morpho-mechanical properties and predictions in Stage I for Pf-RBCs.} Figure descriptions follow Fig.~\ref{fig:7}.} 
\label{fig:8}
\end{figure}
%%%%%%%%%%%%%%%%%%%%%%%%%%%%%%%%%%%%%%%%%%%%%%%%%%%%%%%%%%%%%%%%%%%%%%%%%%%%%%%%%%%%

% hyperparameter
To address the uncertainties mentioned above, a hierarchical Bayesian inference structure is designed in Stage I of our RBC-MsUQ framework to integrate data from different experiments to derive more accurate and experimentally consistent posterior distributions. Based on single-level Bayesian models, an extra level of hyperparameters $\boldsymbol{\psi}_{\mathcal{S}_1}$ is incorporated into the model. These hyperparameters characterize the prior distributions of the model input parameters $\boldsymbol{\vartheta}^{\mathrm{in}}_{\mathcal{S}_1}$ and $\boldsymbol{\sigma}^{\mathrm{out}}_{\mathcal{S}_1}$, thus obtaining collective information from all experimental datasets. Consequently, the selection of hyperparameters strongly influences the effectiveness of the hierarchical Bayesian model. In our RBC-MsUQ framework, for the model input parameters $\boldsymbol{\vartheta}^{\mathrm{in}}_{\mathcal{S}_1}$, the mean and standard deviation of their prior distributions are selected as the corresponding hyperparameters. For example, the hyperparameters for the surface area $A_0$ are its mean $\mu_{A_0}$ and its standard deviation $\sigma_{A_0}$. For the model parameters $\boldsymbol{\sigma}^{\mathrm{out}}_{\mathcal{S}_1}$ that describe the standard deviation of the output values, the lower and upper limits of their range, $L$ and $U$, serve as the hyperparameters. For example, the standard deviation $\sigma_{D_{\mathrm{eq}}}$ of the cell diameter $D_{\mathrm{eq}}$ in the equilibrium tests corresponds to the hyperparameters $L_{\sigma_{D_{\mathrm{eq}}}}$ and $U_{\sigma_{D_{\mathrm{eq}}}}$. Therefore, the hyperparameters in Stage I are made up of two parts: $\boldsymbol{\psi}_{\mathcal{S}_1}=(\boldsymbol{\psi}_{\boldsymbol{\vartheta}^{\mathrm{in}},\mathcal{S}_1},\boldsymbol{\psi}_{\boldsymbol{\sigma}^{\mathrm{out},\mathcal{S}_1}})$, where $\boldsymbol{\psi}_{\boldsymbol{\vartheta}^ {\mathrm{in}},\mathcal{S}_1}=(\mu_{A_0},\sigma_{A_0},\mu_v,\sigma_v,\mu_{\mu_{\mathrm{sh}}},\sigma_{\mu_{\mathrm{sh}}})$, and $\boldsymbol{\psi}_{\boldsymbol{\sigma}^{\mathrm{out}}_{\mathcal{S}_1}}=($$L_{\sigma_{D_{\mathrm{eq}}}},U_{\sigma_{D_{\mathrm{eq}}}},L_{\sigma_{h_{\mathrm{min}}}},U_{\sigma_{h_{\mathrm{min}}}},L_{\sigma_{h_{\mathrm{max}}}},U_{\sigma_{h_{\mathrm{max}}}},L_{\sigma_{D_{\mathrm{ax}}}},U_{\sigma_{D_{\mathrm{ax}}}},L_{\sigma_{D_{\mathrm{tr}}}},U_{\sigma_{D_{\mathrm{tr}}}})$. Based on the distributions of $\boldsymbol{\vartheta}^{\mathrm{in}}_{\mathcal{S}_1}$ and $\boldsymbol{\sigma}^{\mathrm{out}}_{\mathcal{S}_1}$ obtained from single-level Bayesian inference, the posterior distribution of the hyperparameters $p\left(\boldsymbol{\psi} \mid \overrightarrow{\mathbf{d}}, \mathcal{S}_1\right)$ can be inferred, as shown in Fig.~S4 and  Fig.~S5. For both hRBCs and Pf-RBCs, the distributions of hyperparameters have been fully identified on the basis of the results of single-level Bayesian models.

\begin{figure}[!ht]
\begin{center}
\includegraphics[width=1.0\textwidth]{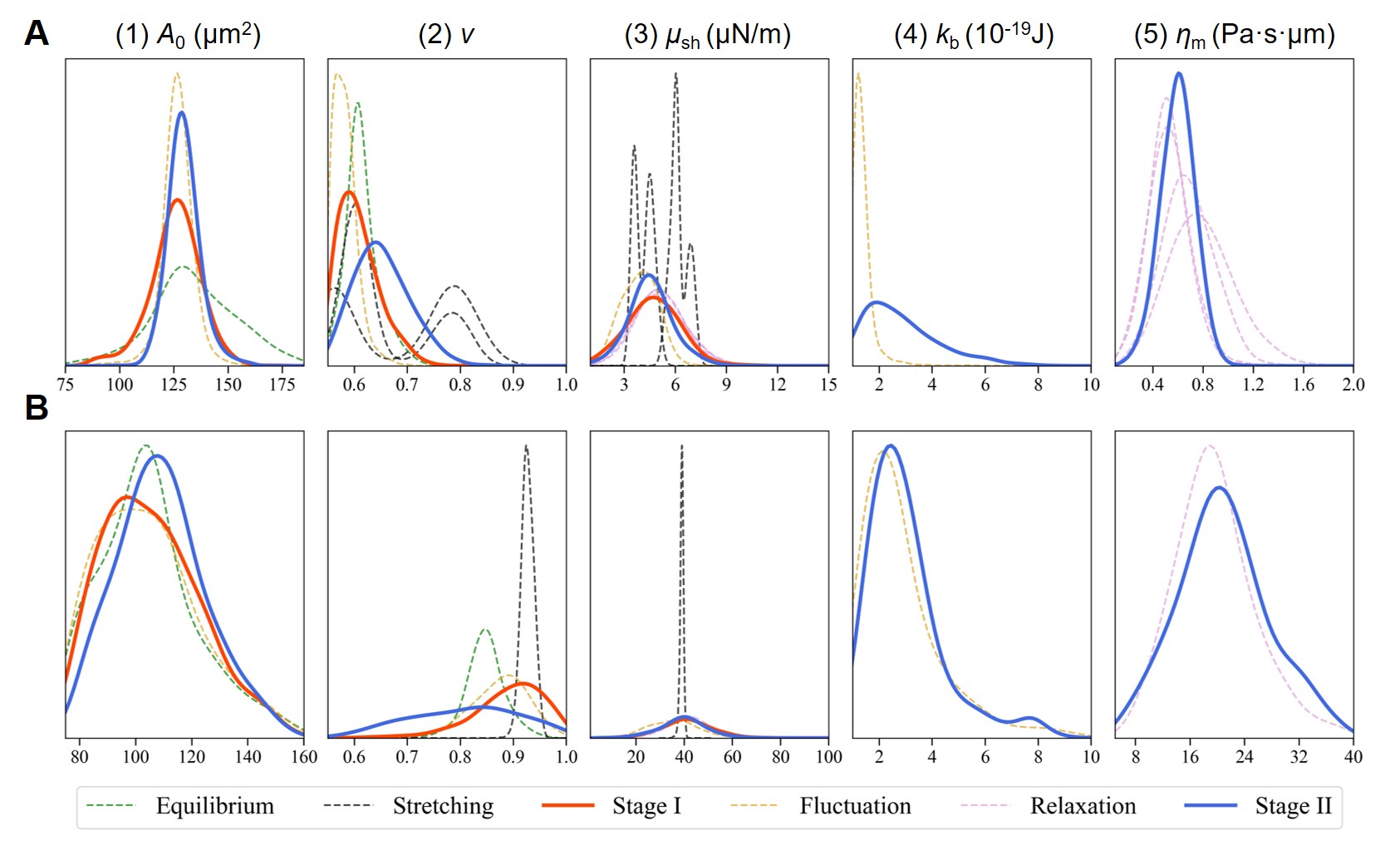}
\end{center}
\vspace{-0.25in}
\caption{\small{\bf Posterior distributions of RBC properties in Stage I and II.} The marginal distributions of the input parameters $\boldsymbol{\vartheta}^{\mathrm{in}}=(A_0,v,\mu_{\mathrm{sh}},k_{\mathrm{b}},\eta_{\mathrm{m}})$ for (A) hRBCs and (B) Pf-RBCs are displayed. The dashed lines in green, black, yellow, and purple represent the posterior distribution from the single-level Bayesian inference based on the equilibrium, stretching, fluctuation, and relaxation tests, whereas the solid lines in red and blue represent the results of hierarchical models in Stage I and II.} 
\label{fig:12}
\end{figure}

% hRBC hierarchical results
After obtaining the posterior distribution of the hyperparameters, forward propagation can be performed to obtain the distribution of the RBC parameter $p\left(\boldsymbol{\vartheta}^{\mathrm{new}} \mid \overrightarrow{\mathbf{d}}, \mathcal{S}_1\right)$, which incorporates data from all experimental datasets, as shown in Fig.~\ref{fig:7}A and Fig.~\ref{fig:8}A. Each parameter exhibits a unimodal distribution with relatively considerable uncertainties around the peaks arising from the heterogeneity between experimental datasets. The predictions of equilibrium and stretching tests $\bm{\mathrm{y}}^\mathrm{new}_{\mathcal{S}_1}$ (dashed lines and shaded regions in Fig.~\ref{fig:7}B-\ref{fig:7}C and Fig.~\ref{fig:8}B-\ref{fig:8}C) based on the posterior distributions exhibit a good concordance with the experimental datasets (black dots). For hRBCs, the posterior $A_{0,\mathcal{S}_1}^{\mathrm{new}}$ (solid red lines in Fig.~\ref{fig:12}A1) is slightly smaller than those obtained from single-level Bayesian inference based on equilibrium tests $A_{0,\mathrm{eq}}^{\mathrm{new}}$ (dashed green lines), mainly due to the coupling changes with reduced volume $v$. The inferred $p\left(v^{\mathrm{new}} \mid \overrightarrow{\mathbf{d}}, \mathcal{S}_1\right)$ (solid red lines in Fig.~\ref{fig:12}A2) with a mean value of 0.60 exhibits a shift to the left compared to $p\left(v^{\mathrm{new}}_1 \mid {\mathbf{d}_i}, \mathcal{M}_{\mathrm{st},i}\right),i=1,2$, resulting from the influence of a prominent distribution near the lower boundary on the results of the stretching tests (dashed black lines). This alteration also leads to the mean value of $y_{h_{\mathrm{max}},\mathcal{S}_1}^{\mathrm{new}}$ for hRBCs being slightly smaller than the experimental data, which still exhibits a high probability in the predicted distribution. Note that the peaks around 0.8 in $p\left(v^{\mathrm{new}}_i \mid {\mathbf{d}_{\mathrm{st},i}}, \mathcal{M}_{\mathrm{st},i}\right),i=1,2$ in the stretching results no longer exist in the posterior distribution $p\left(v^{\mathrm{new}} \mid \overrightarrow{\mathbf{d}}, \mathcal{S}_1\right)$ under the influence of the equilibrium test (if $v=0.8$, the stress-free shape of the RBC will transform into a flat disk instead of a biconcave shape). Therefore, the predicted $y_{D_{\mathrm{tr}},\mathcal{S}_1}^{\mathrm{new}}$ exhibits some deviation from the experimental data at lower stretching forces, while the majority of the experimental values still fall within the 50\% confidence interval, indicating that the model can effectively quantify the model uncertainty drawing from the collective information from multiple experimental datasets. The posterior distribution of the shear modulus $p\left(\mu_{\mathrm{sh}}^{\mathrm{new}} \mid \overrightarrow{\mathbf{d}}, \mathcal{S}_1\right)$ exhibits a well-shaped unimodal form with the mean, median, and ML values being \SI{4.7}{\micro\newton\per\metre}, covering the parameter spaces of the stretching results. The predictions of stretching tests based on the hierarchical model align well with the experimental values that are all included in the wider credible intervals relative to the single-level model, as shown in Fig.~\ref{fig:7}C, demonstrating that the uncertainties quantified in Stage I fully address the diversity of experimental information from different sources.

% Pf-RBC hierarchical results
For Pf-RBCs, the posterior distributions of the model parameters $\boldsymbol{\vartheta}^{\mathrm{in,new}}_{\mathcal{S}_1} =(\boldsymbol{\vartheta}^{\mathrm{in,new}}_{\mathcal{S}_1},\boldsymbol{\sigma}^{\mathrm{out,new}}_{\mathcal{S}_1})$ are also identified in a unimodal form, and the corresponding predictions align well with the experimental values, as shown in Fig.~\ref{fig:8}. Similarly to hRBCs, the distribution of reduced volume $p\left(v^{\mathrm{new}} \mid \overrightarrow{\mathbf{d}}, \mathcal{S}_1\right)$ (solid red lines in Fig.~\ref{fig:12}B2) of Pf-RBCs is updated on the basis of the single-level models of the equilibrium and stretching tests in a coupled manner, with its mean value lying between the single-level results that are encompassed by the broad distribution $p\left(v^{\mathrm{new}} \mid \overrightarrow{\mathbf{d}}, \mathcal{S}_1\right)$ with a larger standard deviation. Upon change in reduced volume, the average values of the prediction $y_{h_{\mathrm{min}},\mathcal{S}_1}^{\mathrm{new}}$ and $y_{h_{\mathrm{min}},\mathcal{S}_1}^{\mathrm{new}}$ are slightly lower than those of the single-level models and experimental measurements, which are still covered by the wide distributions. In general, the distribution of surface area is slightly reduced compared to the single-level results. The shear modulus shows minimal change in its mean value, but its distribution broadens significantly, reflecting the uncertainty introduced by the incorporation of experimental information.

Comparing the results of hRBCs and Pf-RBCs from the hierarchical model, we observe that the surface area of hRBCs is concentrated within the range of 100-\SI{150}{\micro\square\metre}, with a relatively smaller reduced volume that resembles a biconcave shape. In contrast, Pf-RBCs exhibit a loss in surface area and a larger reduced volume, resulting in a shape more akin to an ellipsoid. The shear modulus of hRBCs is concentrated between 1-\SI{8}{\micro\newton\per\metre}, demonstrating excellent deformability, while that of Pf-RBCs is an order of magnitude higher, indicating increased rigidity and difficulty in deformation. These discrepancies are in good agreement with previous experimental observations, illustrating the feasibility of the RBC-MsUQ framework to accurately characterize the morpho-mechanical properties of RBCs in health and disease. In addition to the surface area, reduced volume and shear stress, bending stiffness, and membrane viscosity also play an essential role in the RBC deformation and flow, which are further identified in the subsequent Stage II.

%%%%%%%%%%%%%%%%%%%%%%%%%% Fig9 H2 Model Single Schematic %%%%%%%%%%%%%%%%%%%%%%%%%%
\begin{figure}[!ht]
\begin{center}
\includegraphics[width=0.8\textwidth]{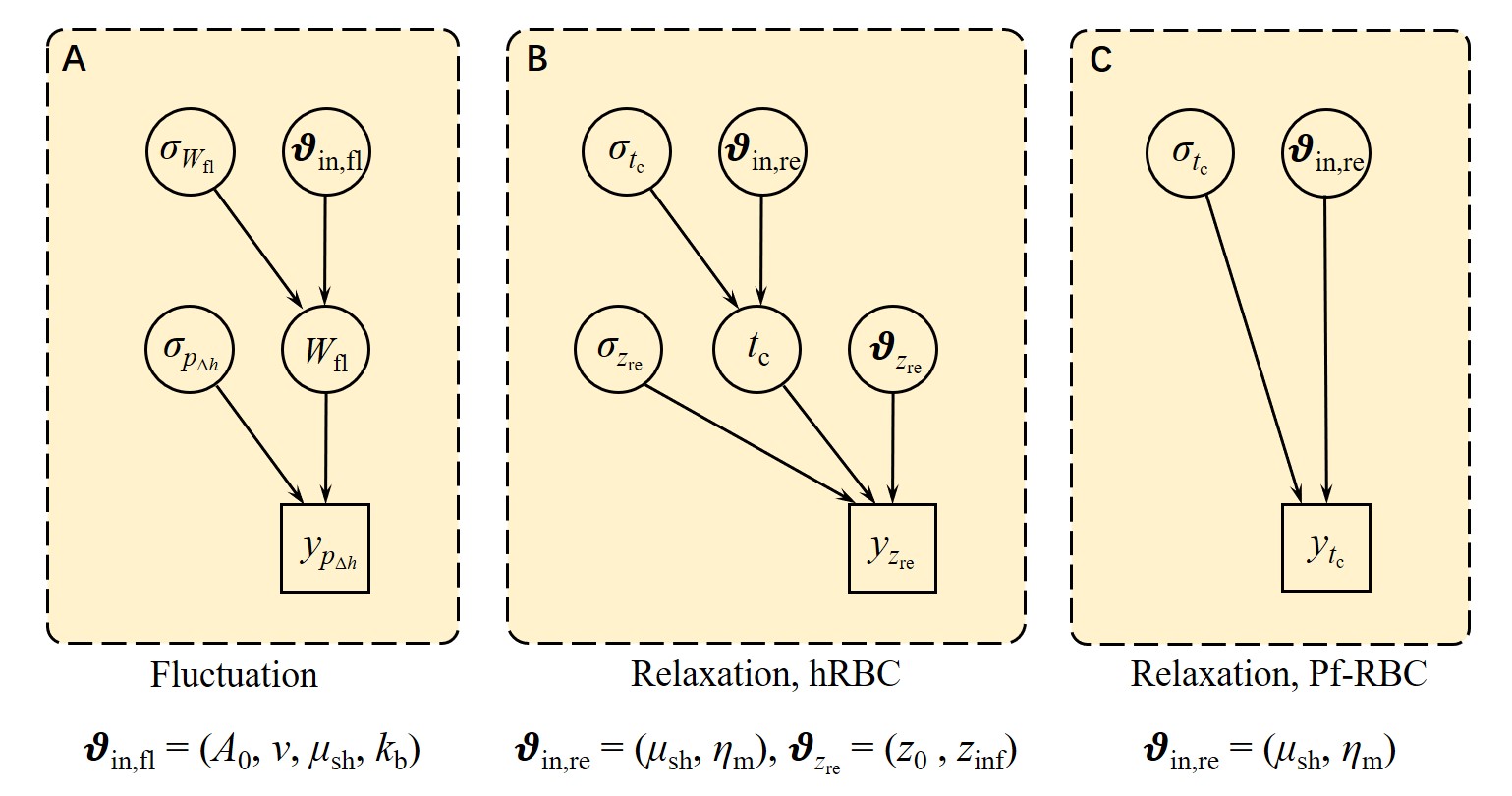}
\end{center}
\vspace{-0.25in}
\caption{\small{\bf Structure of single-level Bayesian model of fluctuation and relaxation tests in Stage II.} (A) Bayesian model of fluctuation test for hRBCs and Pf-RBCs. (B-C) Bayesian model of fluctuation test for (B) hRBCs and (C) Pf-RBCs.} 
\label{fig:9}
\end{figure}
%%%%%%%%%%%%%%%%%%%%%%%%%%%%%%%%%%%%%%%%%%%%%%%%%%%%%%%%%%%%%%%%%%%%%%%%%%%%%%%%%%%%

\subsubsection{Stage II: fluctuation and relaxation tests}
\label{sec_3_3_2}
% fluctuation statistic model
After obtaining the distributions of surface area, reduced volume and shear modulus in Stage I, the hierarchical Bayesian inference in Stage II ($\mathcal{S}_2$) can proceed. First, we conduct single-level Bayesian inference based on the membrane fluctuation when RBCs are adhered to a substrate. However, if we directly use the experimental data, that is, the probability densities $p_{\Delta h}$ of the vertical displacements $\Delta h$, the computational cost for both the neural network and the hierarchical Bayesian inference would be extremely high. Hence, in this work we adopt the full-width half-maximum (denoted as $W_\mathrm{fl}$) of the fluctuation probability distribution curve as the output. The curve is assumed to obey a normal distribution with a mean of zero, and the standard deviation is given by $W_\mathrm{fl}/{2\sqrt{2\ln 2}}$. The structure of the single-level fluctuation Bayesian model is shown in Fig.~\ref{fig:9}A, and the distribution of $W_\mathrm{fl}$ is given by:
\begin{equation}
	W_\mathrm{fl} = G_{W_\mathrm{fl}}(\boldsymbol{\vartheta}^{\mathrm{in}}_{\mathrm{fl}}) + \sigma_{W_\mathrm{fl}} \varepsilon_{W_\mathrm{fl}}\ , \label{eq:RBC-MsUQ-H2-fl1}\\
\end{equation}
where $G_{W_\mathrm{fl}}(\boldsymbol{\vartheta}^{\mathrm{in}}_{\mathrm{fl}})$ is the output $W_\mathrm{fl}$ value of the surrogate model with input of $\boldsymbol{\vartheta}^{\mathrm{in}}_{\mathrm{fl}}$. $\boldsymbol{\vartheta}^{\mathrm{in}}_{\mathrm{fl}}$ consists of surface area $A_0$, reduced volume $v$, shear modulus $\mu_{\mathrm{sh}}$, and bending stiffness $k_{\mathrm{b}}$ based on sensitivity analysis. $\sigma_{W_\mathrm{fl}}$ is the standard deviation of $W_\mathrm{fl}$, with $\varepsilon_{W_\mathrm{fl}} \sim \mathcal{N}(0,1)$. Hence, the statistical model of membrane fluctuation is given by:
\begin{equation}
	y_{p_{\Delta h},1,j} = p_{\Delta h}({\Delta h}_j,W_\mathrm{fl}) + \sigma_{p_{\Delta h},1} \varepsilon_{p_{\Delta h},1}\ , \label{eq:RBC-MsUQ-H2-fl2}\\
\end{equation}
where ${\Delta h}_j,j=1,2,\ldots,n$ corresponds to the membrane displacements in experimental datasets. $\sigma_{p_{\Delta h},1}$ is the standard deviation of $p_{\Delta h}$, with $\varepsilon_{p_{\Delta h}} \sim \mathcal{N}(0,1)$.

%%%%%%%%%%%%%%%%%%%%% Fig10 H2 Model Thetanew Prediction hRBC %%%%%%%%%%%%%%%%%%%%%%
\begin{figure}[!ht]
\begin{center}
\includegraphics[width=1.0\textwidth]{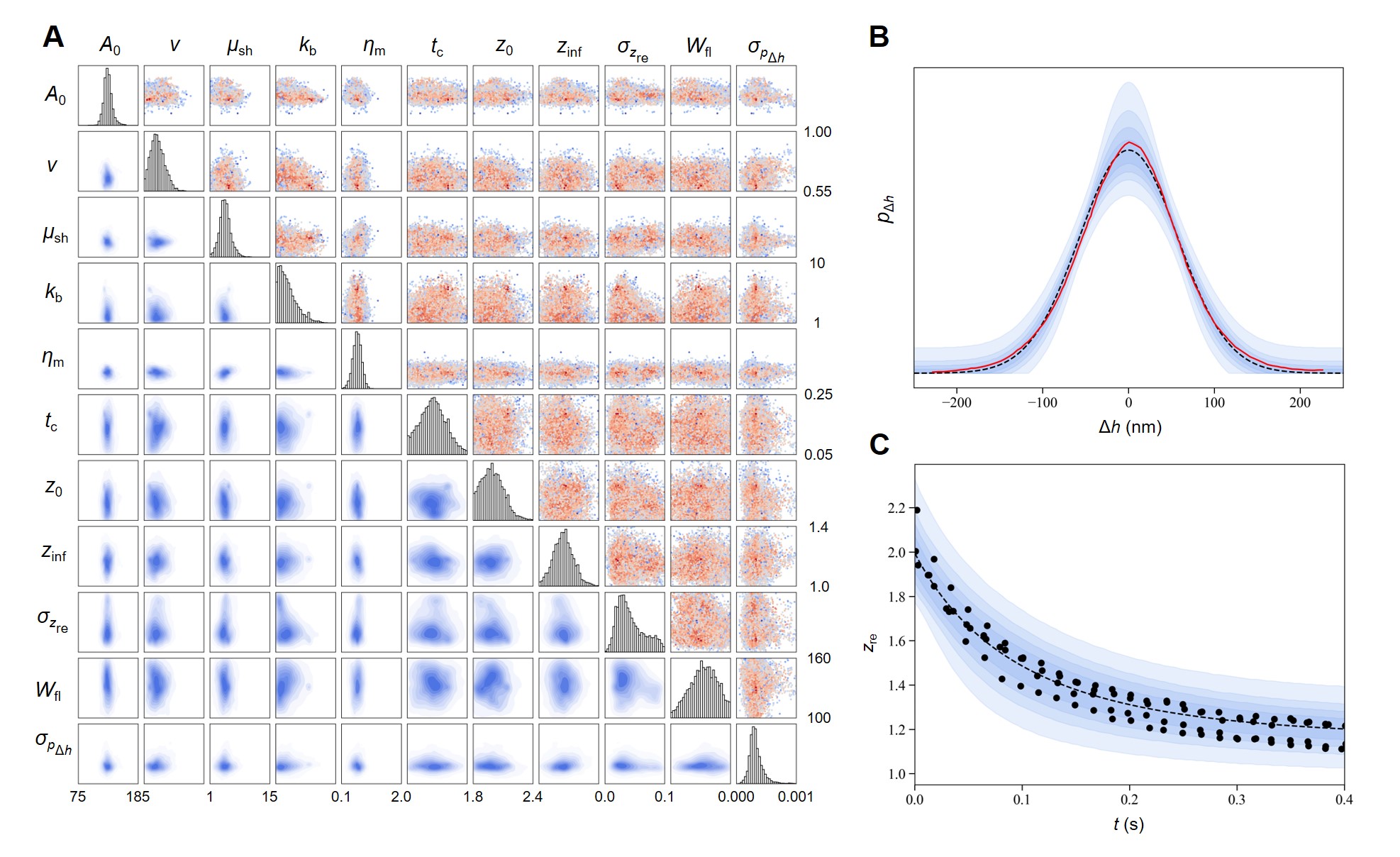}
\end{center}
\vspace{-0.25in}
\caption{\small{\bf Posterior distributions of morpho-mechanical properties and predictions in Stage II for hRBCs.} (A) The posterior distributions of the model parameters $\boldsymbol{\vartheta}_{\mathcal{S}2}$. Figure descriptions follow Fig.~\ref{fig:7}A. (B-C) Predictions derived from the posterior distributions of (B) fluctuation and (C) relaxation tests. The red lines and black dots denote the experimental data, and the dashed lines and shaded regions represent the mean predictions and 50{\%}, 90{\%}, 95{\%}, and 99{\%} credible intervals, respectively.} 
\label{fig:10}
\end{figure}
%%%%%%%%%%%%%%%%%%%%%%%%%%%%%%%%%%%%%%%%%%%%%%%%%%%%%%%%%%%%%%%%%%%%%%%%%%%%%%%%%%%%

Since membrane fluctuation is sensitive to four parameters $\boldsymbol{\vartheta}^{\mathrm{in}}_{\mathrm{fl}}=(A_0,v,\mu_{\mathrm{sh}},k_{\mathrm{b}})$, it is difficult to obtain stable distributions of these parameters from a single type of experimental result. Therefore, the distribution information of surface area $A_0$, reduced volume $v$, and shear modulus $\mu_{\mathrm{sh}}$ obtained in Stage I is transferred to Stage II, ensuring that the distribution updated through the single-level fluctuation Bayesian model also aligns with the results from Stage I, thus yielding stable distributions. As the input parameter samples are continuously updated during the TMCMC sampling process, it is challenging to directly sample the posterior samples of Stage I. Consequently, the posterior distribution of hyperparameters from Stage I is transmitted to Stage II. When performing single-level Bayesian inference, forward propagation from the hyperparameters to the posterior distribution is incorporated into the inference process. Through this approach, we further obtain stable results of single-level Bayesian inference for fluctuation tests, as shown in Fig.~S6.

% fluctuation single results
The posterior distributions of $\boldsymbol{\vartheta}_\mathrm{fl}$ in the single-level Bayesian model for fluctuation tests are well updated to a unimodal form, as shown in Figs.~S6A-S6B. The values of the mean, median and maximum likelihood estimates of surface area $A_0$, reduced volume $v$, and shear modulus $\mu_{\mathrm{sh}}$ are similar to the results of Stage I, indicating that the parameter information of Stage I has been effectively transmitted to the single-level Bayesian inference process in Stage II. It is worth noting that the distribution of the shear modulus $\mu_{\mathrm{sh}}$ is slightly reduced compared to the posterior distribution results in Stage I, whether for hRBCs or Pf-RBCs. This discrepancy originates from the fact that $\mu_{\mathrm{sh}}$ is identifiable not only from the experimental datasets in Stage I, but also from the fluctuation tests in Stage II, which have been revealed through the sensitivity analysis. The distributions of bending stiffness $k_{\mathrm{b}}$ are also updated to narrow peaks around 1.53$\times 10^{-19}$\SI{}{\J} for hRBCs, which lies within previous experimental observations and slightly lower than the commonly used value 2.4$\times 10^{-19}$\SI{}{\J} in previous viscoelastic models for RBCs. The posterior $k_{\mathrm{b}}$ for Pf-RBCs is slightly higher than that for hRBCs, but the difference is not as significant as other morpho-mechanical properties, such as shape or shear modulus. This finding is consistent with the lack of a unified conclusion in previous experiments on changes in bending rigidity during malaria infection. The predictions based on the posterior distribution (black dashed lines in Fig.~S6C-S6D) align well with the experimental values (red solid lines), indicating that the experimental information has been effectively used to infer the RBC parameter distribution. However, because of the difference between the membrane displacement distribution curve of Pf-RBCs and the Gaussian distribution, the confidence interval for the prediction of Pf-RBCs is relatively larger than that of hRBCs. This reflects the uncertainty generated by the discrepancy between the model and the experimental data.

% relaxation statistic model
Subsequently, the membrane viscosity $\eta_{\mathrm{m}}$ along with the shear modulus $\mu_{\mathrm{sh}}$ are updated through relaxation tests. Since the experimental results for hRBCs include changes in cell deformation $z_{\mathrm{re}}$ over time $t$, while the form of the results for Pf-RBCs corresponds to the relaxation time $t_{\mathrm{c}}$ (see Table~\ref{table:Exp}), different statistical models are employed for hRBCs and Pf-RBCs, as shown in Fig.~\ref{fig:9}B-\ref{fig:9}C.

For hRBCs, the experimental datasets for the relaxation model consist of the changes in the ratio $z_{\mathrm{re}}$ between axial length $D_{\mathrm{ax}}$ and transverse length $D_{\mathrm{tr}}$ over time $t$ upon the stretching release from four sets of relaxation tests. However, similar to the fluctuation model, directly using $z_{\mathrm{re}}$ as the output of the neural network and hierarchical Bayesian inference would require significant computational costs. Therefore, it is assumed that the relaxation curve follows the exponential decay model proposed by Hochmuth~\cite{hochmuth1979red}, as described in the following equation:
\begin{equation}
    z_{\mathrm{re}}(t,t_{\mathrm{c}})=z_{\infty}\frac{\Lambda+e^{-t/t_\mathrm{c}}}{\Lambda-e^{-t/t_\mathrm{c}}}, \label{eq:RBC-MsUQ-H2-z}\\
\end{equation}
where $\Lambda$ is expressed by:
\begin{equation}
    \Lambda=\frac{z_{0}+z_{\infty}}{z_{0}-z_{\infty}}, \label{eq:RBC-MsUQ-H2-Lamba}\\
\end{equation}
where the characteristic relaxation time $t_{\mathrm{c}}$ describes the relaxation characteristics of RBCs, while $z_{0}$ and $z_{\infty}$ denote the initial and final values of $z_{\mathrm{re}}$ during the relaxation phase, respectively. In the relaxation Bayesian model for hRBCs, $t_{\mathrm{c}}$ is used as the output quantity of the surrogate model instead of $z_{\mathrm{re}}$, and it also aligns with $\boldsymbol{\vartheta}_{z_\mathrm{re}}=(z_{0},z_{\infty})$ to obtain the fitted curve $z_{\mathrm{re}}(t,t_{\mathrm{c}})$. The characteristic relaxation time $t_{\mathrm{c}}$ is given by:
\begin{equation}
    t_{\mathrm{c}} = G_{t_{\mathrm{c}}}(\boldsymbol{\vartheta}^{\mathrm{in}}_{\mathrm{re}}) + \sigma_{t_{\mathrm{c}}} \varepsilon_{t_{\mathrm{c}}}\ , \label{eq:RBC-MsUQ-H2-re1}\\
\end{equation}
where $G_{t_{\mathrm{c}}}(\boldsymbol{\vartheta}^{\mathrm{in}}_{\mathrm{re}})$ represents the output of the surrogate model when the input morpho-mechanical parameters are $\boldsymbol{\vartheta}^{\mathrm{in}}_{\mathrm{re}}=(\mu_{\mathrm{sh}},\eta_{\mathrm{m}})$. $\sigma_{t_{\mathrm{c}}}$ is the standard deviation of $t_{\mathrm{c}}$, and $\varepsilon_{t_{\mathrm{c}}} \sim \mathcal{N}(0,1)$. Thus, the statistical equation for the relaxation test is given by:
\begin{equation}
    y_{z_{\mathrm{re}},i,j} = z_{\mathrm{re}}(t_j,t_{\mathrm{c}},\boldsymbol{\vartheta}_{z_\mathrm{re}}) + \sigma_{z_{\mathrm{re}},i} \varepsilon_{z_{\mathrm{re}},i}\ , \label{eq:RBC-MsUQ-H2-re2}\\
\end{equation}
where $t_j,j=1,2,\ldots,n$ are the displacement values obtained from the experiment, and the subscript $i$ denotes the index of the experimental datasets. For hRBCs, $i=1,2,3,4$. $\sigma_{z_{\mathrm{re}},i}$ and $\varepsilon_{z_{\mathrm{re}}} \sim \mathcal{N}(0,1)$ are the standard deviation and independent Gaussian noise for $z_{\mathrm{re}}$, respectively.

In contrast to hRBCs, previous experimental data for Pf-RBCs lack the measurement of the time-dependent deformation during the relaxation process. Hence, the statistical model is given by:
\begin{equation}
    y_{t_{\mathrm{c}},1} = G_{t_{\mathrm{c}}}(\boldsymbol{\vartheta}^{\mathrm{in}}_{\mathrm{re}}) + \sigma_{t_{\mathrm{c}},1} \varepsilon_{t_{\mathrm{c}},1}\ , \label{eq:RBC-MsUQ-H2-re3}\\
\end{equation}
where $G_{t_{\mathrm{c}}}(\boldsymbol{\vartheta}^{\mathrm{in}}_{\mathrm{re}})$ represents the surrogate model for the relaxation test of Pf-RBCs.

Similarly to the fluctuation tests, single-level Bayesian inference performed individually on the relaxation test cannot generate stable results. Although the output of the relaxation test is merely dependent on the shear modulus $\mu_{\mathrm{sh}}$ and the membrane viscosity $\eta_{\mathrm{m}}$, previous experiments have revealed that the characteristic relaxation time $t_{\mathrm{c}}$ is approximately the ratio of membrane viscosity to shear modulus, i.e. $t_{\mathrm{c}} \sim \eta_{\mathrm{m}}/\mu_{\mathrm{sh}}$~\cite{hochmuth1979red}. Therefore, single-level Bayesian inference based on the relaxation experiment data sets can lead to a stable distribution of $\eta_{\mathrm{m}}/\mu_{\mathrm{sh}}$, while the individual distributions of the two input parameters remain unstable. As a result, the hyperparameter posterior distributions obtained in Stage I are also transferred to the relaxation test, and the forward propagation from the hyperparameters to input parameters is incorporated, allowing for stable results in the single-level Bayesian inference of the relaxation test.

%%%%%%%%%%%%%%%%%%%%% Fig11 H2 Model Thetanew Prediction Pf-RBC %%%%%%%%%%%%%%%%%%%%
\begin{figure}[!ht]
\begin{center}
\includegraphics[width=1.0\textwidth]{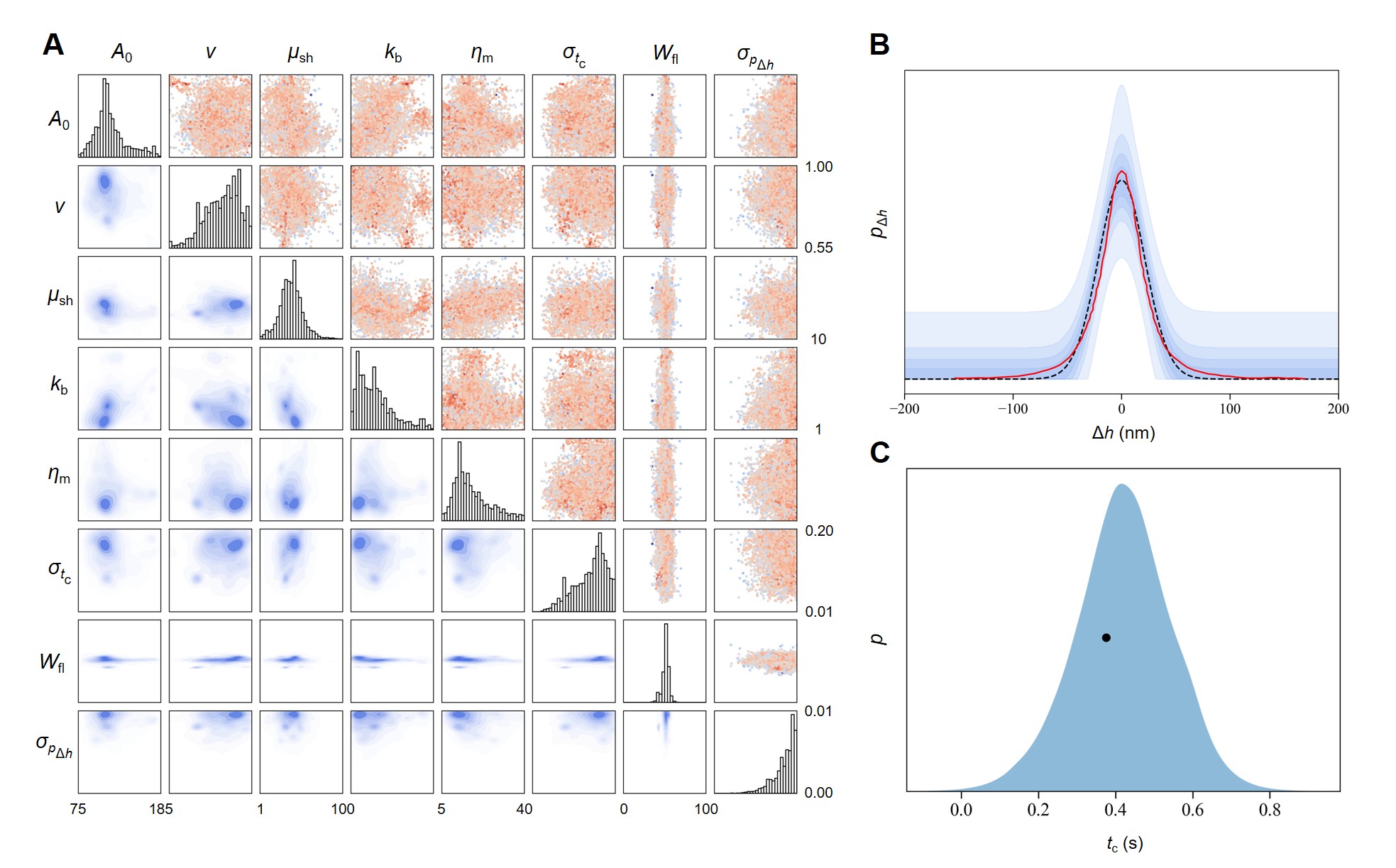}
\end{center}
\vspace{-0.25in}
\caption{\small{\bf Posterior distributions of morpho-mechanical properties and predictions in Stage II for Pf-RBCs.} Figure descriptions follow Fig.~\ref{fig:10}.} 
\label{fig:11}
\end{figure}
%%%%%%%%%%%%%%%%%%%%%%%%%%%%%%%%%%%%%%%%%%%%%%%%%%%%%%%%%%%%%%%%%%%%%%%%%%%%%%%%%%%%

% relaxation single results
The results of the single-level Bayesian inference of the relaxation test are shown in Fig.~S7. The four columns on the left represent results for hRBCs, each obtained from one of four sets of relaxation experimental data, while the right column represents the results for Pf-RBCs. Due to the distribution information of the hyperparameters from Stage I, the distributions of the shear modulus $\mu_{\mathrm{sh}}$ are similar to the posterior distribution in Stage I. For hRBCs, the mean values of $\mu_{\mathrm{sh}}$ based on four datasets are concentrated between 4.8 and 5.2 \SI{}{\micro\newton\per\metre}, while for Pf-RBCs, the average is 42.56 \SI{}{\micro\newton\per\metre} (see Table~S1). The membrane viscosity distributions $\eta_{\mathrm{m}}$ for both hRBCs and Pf-RBCs are also fully identified from the experimental information. For hRBCs, the mean values of $\eta_{\mathrm{m}}$ from the four datasets of relaxation experiments are 0.52, 0.53, 0.68, and 0.79 \SI{}{\Pa\cdot\s\cdot\micro\metre}, respectively, which are higher than the commonly used value of 0.38 \SI{}{\Pa\cdot\s\cdot\micro\metre} in previous computational models. There are marked discrepancies between the four distributions, reflecting the inherent uncertainties in the experimental results. For Pf-RBCs, the mean, median and maximum likelihood estimates of $\eta_{\mathrm{m}}$ are 19.67, 19.21, and 17.49 \SI{}{\Pa\cdot\s\cdot\micro\metre}, which are an order of magnitude higher than those of hRBCs. This indicates a significant alteration in membrane properties during parasitization by the malaria parasite, and cell deformation will take much longer than that of hRBCs.
In the bottom row of Fig.~S7, the predicted values (blue area and black dashed line) obtained by incorporating the posterior distribution into the surrogate model are compared with the experimental values (black points). For both hRBCs and Pf-RBCs, each set of predicted values aligns well with the corresponding experimental data, demonstrating the effectiveness of single-level Bayesian inference in parameter estimation based on individual experimental datasets.

% hyperparameter
To further integrate the information from different datasets and Stage I, a hierarchical structure is also essential in Stage II. Similarly to stage I, a level of hyperparameters $\boldsymbol{\psi}_{\mathcal{S}_2}$ is constructed that governs the prior distributions of the model parameters. The mean and SD values of the input parameters $\boldsymbol{\vartheta}^{\mathrm{in}}_{\mathcal{S}_2}=(A_0,v,\mu_{\mathrm{sh}},k_{\mathrm{b}},\eta_{\mathrm{m}})$, and the upper and lower limits of the SDs of the output quantities $\boldsymbol{\sigma}^{\mathrm{out}}_{\mathcal{S}_2}$ are adopted as hyperparameters. For hRBCs, $\boldsymbol{\sigma}^{\mathrm{out}}_{\mathcal{S}_2}=(\sigma_{p_{\Delta h}},\sigma_{z_{\mathrm{re}}})$, while for Pf-RBCs, $\boldsymbol{\sigma}^{\mathrm{out}}_{\mathcal{S}_2}=(\sigma_{p_{\Delta h}},\sigma_{t_{\mathrm{c}}})$. Unlike stage I, intermediate variables including $W_\mathrm{fl}$ and $t_{\mathrm{c}}$ are introduced into the relaxation and fluctuation models to save computational cost; hence, their mean and SD are also included in the hyperparameters. With a level of these hyperparameters, the hierarchical Bayesian inference in Stage II can be performed, and the posterior distributions of hyperparameters $p\left(\boldsymbol{\psi} \mid \overrightarrow{\mathbf{d}}, \mathcal{S}_2\right)$ for hRBCs and Pf-RBCs are identified, as shown in Fig.~S8 and Fig.~S9.

%%%%%%%%%%%%%%%%%%%%%%%%%%%%%% Fig12 All distributions %%%%%%%%%%%%%%%%%%%%%%%%%%%%%
%%%%%%%%%%%%%%%%%%%%%%%%%%%%%%%%%%%%%%%%%%%%%%%%%%%%%%%%%%%%%%%%%%%%%%%%%%%%%%%%%%%%

After inferring the posterior distribution of the hyperparameters, forward propagation is conducted to obtain the posterior distribution $p\left(\boldsymbol{\vartheta}^{\mathrm{new}} \mid \overrightarrow{\mathbf{d}}, \mathcal{S}_2\right)$ of the model parameters, as shown in Fig.~\ref{fig:10}A and  Fig.~\ref{fig:11}A. The distributions of input parameters that have been inferred in Stage I, including surface area $A_{0,{\mathcal{S}_2}}^{\mathrm{new}}$, reduced volume $v_{{\mathcal{S}_2}}^{\mathrm{new}}$ and shear modulus $\mu_{\mathrm{sh},{\mathcal{S}_2}}^{\mathrm{new}}$, are similar to the results in Stage I, indicating the feasibility of the strategy that transfers hyperparameters from Stage I to Stage II. For hRBCs, the mean value of bending stiffness $k_{\mathrm{b},{\mathcal{S}_2}}^{\mathrm{new}}$ is 2.84$\times 10^{-19}$\SI{}{\J}, close to the commonly used value of 2.4$\times 10^{-19}$\SI{}{\J} in previous computational models. Meanwhile, the inferred $k_{\mathrm{b},{\mathcal{S}_2}}^{\mathrm{new}}$ for Pf-RBCs has a mean of 3.02$\times 10^{-19}$\SI{}{\J}, which is similar to hRBCs, indicating that the impact of Plasmodium parasitization on the bending rigidity of red blood cells is minimal. However, there exists a substantial difference in the posterior viscosity distributions between hRBCs and Pf-RBCs, reflecting the high-viscosity property of Pf-RBCs.

The comparison between the predicted values and experimental datasets presented in Fig.~\ref{fig:10}B-\ref{fig:10}C and  Fig.~\ref{fig:11}B-\ref{fig:11}C demonstrates that for both hRBCs and Pf-RBCs, the predictions align well with the experimental datasets. Compared to the predictions from the single-level Bayesian inference, the standard deviations of the predictions based on the hierarchical model are generally larger, and they may not perfectly match the experimental values. For instance, the predicted mean value of the characteristic relaxation time $t_{\mathrm{c}}$ for Pf-RBCs is slightly higher than the experimental value, while the single-level results are closer to the experimental data. However, this originates from the fact that hierarchical Bayesian inference integrates information from multiple sets of experimental data, yielding a more comprehensive understanding of the RBC morpho-mechanics. These posterior distributions account for the uncertainties of different resources, and the resulting parameters are more representative of the actual distribution of red blood cell parameters.

\subsubsection{Validation of the RBC-MsUQ framework through slit retention test}
\label{sec_3_4}
%%%%%%%%%%%%%%%%%%%%%%%%%%%%%%% Fig13 Slit retention %%%%%%%%%%%%%%%%%%%%%%%%%%%%%%%
\begin{figure}[!tbp]
\begin{center}
\includegraphics[width=1.0\textwidth]{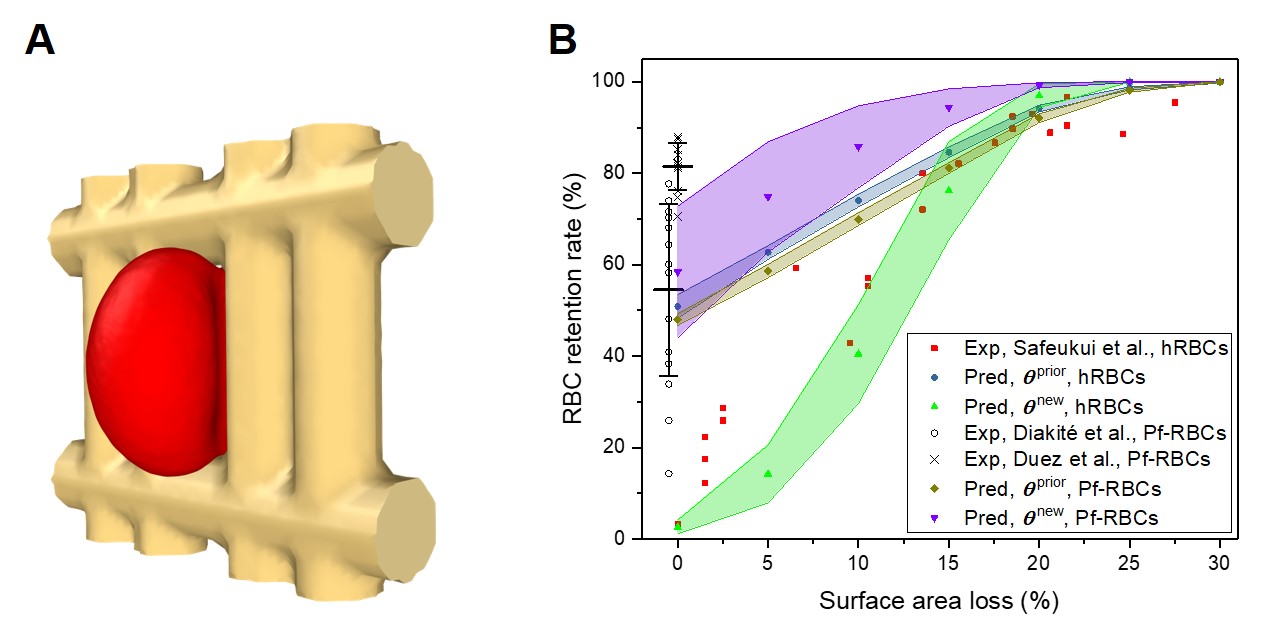}
\end{center}
\vspace{-0.25in}
\caption{\small{\bf Validation of the RBC-MsUQ framework through slit retention experiments.} (A) Diagram illustrating the simulation setup for slit retention experiments, where the RBC, represented in red, with specific morpho-mechanical properties is subjected to pressure as it traverses the narrow slit. (B) Slit retention rates of RBCs are shown for the prior distribution of hRBCs (light blue) and Pf-RBCs (light brown), and the posterior distribution of hRBCs (light green) and Pf-RBCs (purple), and the shaded regions refer to the corresponding error bars. Red data points represent experimental results for hRBCs as reported by Safeukui et al.~\cite{safeukui2012quantitative}. In contrast, black circles and crosses indicate the experimental results for Pf-RBCs during the ring and schizont stages, respectively. The corresponding means and SDs are also provided, as referenced in the studies by Diakit\'{e} et al.~\cite{diakite2016stage} and Duez et al.~\cite{duez2015splenic}. For clarity, the data points have been shifted horizontally.} 
\label{fig:13}
\end{figure}
%%%%%%%%%%%%%%%%%%%%%%%%%%%%%%%%%%%%%%%%%%%%%%%%%%%%%%%%%%%%%%%%%%%%%%%%%%%%%%%%%%%%

The predictions from the four representative mechanical experiments, when compared to the experimental datasets, demonstrate the accuracy of the model's posterior distribution. To further evaluate the generalizability of the model results, we introduce a test that involves the passage of RBCs through a narrow slit, as illustrated in Fig.~\ref{fig:13}. The blockage of RBCs in the splenic slit is a critical and life-threatening consequence of malaria, significantly influenced by the morpho-mechanical properties of RBCs. The retention rate serves as a crucial benchmark for assessing RBC behavior; however, in previous {\em{ in silica}} studies, this rate has rarely been determined due to the lack of distribution data regarding RBC properties. When single parameters are considered in isolation, computational models tend to yield binary outcomes---either passage or blockage---failing to accurately capture the retention proportions. 

In this study, we address this issue using the posterior distributions derived from our RBC-MsUQ framework. Following the {\em{ex vivo}} experiments conducted by Safeukui et al.~\cite{safeukui2012quantitative}, we perform corresponding simulations in which the RBCs traverse a narrow splenic slit, as shown in Fig.~\ref{fig:13}A. Based on previous computational simulations of RBC traversal through slits,~\cite{pivkin2016biomechanics,ma2023computational,qi2021quantitative} we set the slit dimensions to \SI{4.0}{\micro\metre}$\times$\SI{1.2}{\micro\metre}, applying a driving pressure of 1 Pa to facilitate the passage of the RBCs. 
%We impose a reflection boundary condition on the slit walls to ensure complete contact and prevent overlapping between the RBCs and the slit~\cite{ma2023computational}. 
Our slit retention simulations employ RBC properties that conform to both the prior and posterior distributions of hRBCs and Pf-RBCs, incorporating surface area loss for comparison with experimental results. The retention rates derived from the four distributions, along with the experimental datasets for hRBCs (red points) and Pf-RBCs (black circles and crosses), are presented in Fig.~\ref{fig:13}B. The light blue, light brown, light green, and purple points, along with their respective shaded regions, represent the predicted retention rates and the corresponding error bars for the prior and posterior distributions of hRBCs and Pf-RBCs. 

The predictive results derived from the posterior distribution of hRBCs align closely with the experimental findings, thereby corroborating the accuracy of our inferential framework. In contrast, predictions based on the prior distribution of hRBCs significantly overestimate the retention rate, revealing a discrepancy between the assumed parameter set $\boldsymbol{\vartheta}^{\mathrm{prior}}$ and the actual physiological values. For Pf-RBCs, the absence of experimental data linking retention rates to extra surface area loss necessitates a focus on existing datasets that do not include this variable for the ring stage (black circles) and the schizont stage (black crosses)~\cite{diakite2016stage,duez2015splenic}. The predictions based on prior distributions yield a mean retention rate of approximately 48.0\% with relatively narrow SDs, while estimates derived from the posterior distribution start around 58.4\%, featuring broader distributions, and approach 100.0\% retention as surface area loss increases. This trend is consistent with experimental observations, which indicate that retention rates increase from approximately 54.5\%~\cite{diakite2016stage} to 81.5\%~\cite{duez2015splenic} as Pf-RBCs mature from the ring to the schizont stages, coinciding with reductions in surface area and increases in sphericity---both of which are known to enhance RBC retention~\cite{safeukui2013surface}. Although the quantitative comparison between model predictions and experimental measurements is constrained by limited experimental metrics, the results unequivocally demonstrate that Pf-RBCs exhibit significantly higher retention rates than hRBCs, emphasizing their impaired deformability and compromised microcirculatory function. %Through the prediction and validation processes employed in our RBC-MsUQ framework, we demonstrate both the accuracy and robustness of our identified parameter distribution.

%%%%%%%%%%%%%%%%%%%%%%%%%%%%%%%%%%%%%%%%%%%%%%%%%%%%%%%%%%%%%%%%%%%%%%%%%%%%%%%%%%%%%%%%%%%%%%%%%%%%%%%%%%%%%%%%%%%%%%

\section{Conclusion}

In this study, we introduce RBC-MsUQ, a multi-stage uncertainty quantification (UQ) framework for red blood cell (RBC) morpho-mechanical properties that integrates cross-platform experimental datasets via hierarchical Bayesian inference. This approach enables the estimation of posterior distributions for the morpho-mechanical properties of healthy RBCs (hRBCs) and Plasmodium falciparum-infected RBCs (Pf-RBCs), validated through systematic comparisons between model predictions and experimental observations. 

Five critical morpho-mechanical parameters that govern the behaviors of RBCs are identified: surface area ($A_0$), reduced volume ($v$), shear stiffness ($\mu_{\mathrm{sh}}$), bending stiffness ($k_{\mathrm{b}}$), and membrane viscosity ($\eta_{\mathrm{m}}$), treated as unobserved variables in the model. The prior distributions for these parameters are derived from microscopic simulations and prior experimental studies, which are subsequently refined through preliminary Bayesian inference. The RBC-MsUQ framework encompasses simulations of four representative mechanical experiments---equilibrium state, stretching test, membrane fluctuation, and relaxation test---initiated with a uniform prior distribution. To establish a stress-free baseline essential for RBC morpho-mechanical characterization, we implement a dynamic annealing technique to relieve surface stress, yielding the appropriate RBC morphology, particularly under conditions of elevated reduced volume. 

Following rigorous simulations, surrogate neural network models are developed for the hRBC and Pf-RBC experiments, achieving an error margin of the order of $10^{-2}$. These surrogate models facilitate sensitivity analysis of output variables to the input parameters, enhancing computational efficiency in subsequent inference stages. Hierarchical Bayesian inference models are constructed in a two-stage architecture, integrating experimental data to iteratively refine parameter distributions. The resulting posterior distributions reveal that Pf-RBCs exhibit increased stiffness, heightened membrane viscosity, decreased surface area, and a more spherical morphology compared to hRBCs. Model predictions based on posterior distributions show agreement with experimental datasets, validating the accuracy of the RBC-MsUQ framework. 
%Finally, the slit retention tests are conducted and the results of posterior distribution of hRBCs is in consistence with Safeukui's {\em{ex vivo}} experiments, while the Pf-RBCs exhibit high retention rate due to the structural change.

In summary, this research presents a generalizable multi-stage framework for elucidating the distribution of cell morpho-mechanical properties under defined experimental conditions. The RBC-MsUQ framework is readily applicable to other hematological disorders (e.g. diabetes, sickle cell anemia) for systematic investigation of RBC alterations in pathological scenarios. Beyond the four representative assays herein, the methodology enables integration of additional quantitative experimental approaches (e.g. micropipette aspiration, microfiltration, and magnetic twisting cytometry) and computational models (e.g. smoothed particle hydrodynamics, lattice Boltzmann method). Furthermore, coupling with physics-informed neural networks (PINN)~\cite{raissi2019physics,lin2022multi} and deep operator networks (DeepONet)~\cite{lu2021learning} could enhance the resilience of the RBC-MsUQ framework to data noise, model-experiment discrepancies, and surrogate model uncertainties~\cite{zou2024neuraluq,zou2025uncertainty}. This extensible framework thus serves as a robust tool for systematic and quantitative characterization of morpho-mechanical properties in both natural and bioengineered cellular systems.

%%%%%%%%%%%%%%%%%%%%%%%%%%%%%%%%%%%%%%%%%%%%%%%%%%%%%%%%%%%%%%%%%%%%%%%%%%%%%%%%%%%%%%%%%%%%%%%%%%%%%%%%%%%%%%%%%%%%%%
\section*{Acknowledgments}

This work was supported by the National Key R\&D Program (No. 2022YFA1203202), the National Key R\&D Plan ``Inter-governmental International Science \& Technology Innovation Cooperation" Key Specialized Program (No. 2025YFE0107500) and the National Natural Science Foundation of China (No. 12372265). Simulations were conducted at the Beijing Super Cloud Computing Center (BLSC).

\section*{Declaration of interest}

The authors declare that they have no known competing financial interests or personal relationships that could have appeared to influence the work reported in this paper.

\section*{Data availability}

Data will be made available on request.

\section*{CRediT authorship contribution statement}

\textbf{Shuo Wang:} Methodology, Software, Investigation, Data curation, Writing - original draft. \textbf{Lei Ma:} Methodology, Investigation, Writing - review \& editing. \textbf{Ling Guo:} Conceptualization, Methodology, Supervision, Writing- review \& editing. \textbf{Xuejin Li:} Conceptualization, Methodology, Supervision, Writing- review \& editing. \textbf{Tao Zhou:} Supervision, Writing- review \& editing. 

\bibliographystyle{elsarticle-num}
\bibliography{ref}
\biboptions{sort&compress}

\end{document}